\documentclass[10pt,prb,twocolumn]{revtex4}
\usepackage{graphicx,amsmath,amssymb,slashed,dcolumn,bm,color}

\begin{document}
\newcommand{\addRN}[1]{\textcolor{blue}{#1}}
\newcommand{\addJM}[1]{\textcolor{red}{#1}}
\newcommand{\be}{\begin{equation}}
\newcommand{\ee}{\end{equation}}
\newcommand{\bea}{\begin{eqnarray}}
\newcommand{\eea}{\end{eqnarray}}
\newcommand{\HH}{{\cal H}}
\newcommand{\LL}{{\cal L}}
\newcommand{\KK}{{\cal K}}
\newcommand{\VV}{{\cal V}}
\newcommand{\GG}{{\sf G}}
\newcommand{\tr}{{\rm tr\/}\,}
\newcommand{\p}{\partial}
\newcommand{\s}{\sigma}
\newcommand{\la}{\langle}
\newcommand{\ra}{\rangle}
\newcommand{\lb}{\left[}
\newcommand{\rb}{\right]}
\newcommand{\lp}{\left(}
\newcommand{\rp}{\right)}
\newcommand{\E}{{\cal E}}
\newcommand{\der}{{\partial}}
\newcommand{\Tr}{{\rm Tr}\,}
\renewcommand{\vec}[1]{{\bf #1}}
\renewcommand{\Im}{{\rm Im}\,}
\renewcommand{\Re}{{\rm Re}\,}
\def\nn{\nonumber\\}
\newcommand{\mpar}[1]{\marginpar{\small \it #1}}
\newcommand{\diag}{\mathrm{diag}}
\newcommand{\sign}{\mathrm{sign}}

\title{Superconductivity of disordered Dirac fermions}

\author{Rahul Nandkishore$^1$, Joseph Maciejko$^1$, David A. Huse$^{1,2}$, and S. L. Sondhi$^2$}

\affiliation{$^1$Princeton Center for Theoretical Science, Princeton University, Princeton, New Jersey 08544, USA\\ $^2$Department of Physics, Princeton University, Princeton, New Jersey 08544, USA
}
 
 \date\today
 
\begin{abstract}
We study the effect of disorder on massless, spinful Dirac fermions in two spatial dimensions with attractive interactions, and show that the combination of disorder and attractive interactions is deadly to the Dirac semimetal phase. First, we derive the zero temperature phase diagram of a clean Dirac fermion system with tunable doping level ($\mu$) and attraction strength ($g$). We show that it contains two phases: a superconductor and a Dirac semimetal. Then, we add disorder, and show that arbitrarily weak disorder destroys the Dirac semimetal, turning it into a superconductor instead. Thus for Dirac fermions near charge neutrality, disorder actually assists superconductivity. We discuss the strength of the superconductivity for both long range and short range disorder. For long range disorder, the superconductivity is exponentially weak in the disorder strength. For short range disorder, a uniform mean field analysis predicts that superconductivity should be doubly exponentially weak in the disorder strength. However, a more careful treatment of mesoscopic fluctuations suggests that locally superconducting puddles should form at a much higher temperature, and should establish global phase coherence at a temperature that is only exponentially small in weak disorder. Thus, mesoscopic fluctuations exponentially enhance the superconducting critical temperature. We also discuss the effect of disorder on the quantum critical point of the clean system, building in the effect of disorder through a replica field theory. We show that disorder is a relevant perturbation to the supersymmetric quantum critical point. We expect that in the presence of attractive interactions, the flow away from the critical point ends up in the superconducting phase, although firm conclusions cannot be drawn since the renormalization group analysis flows to strong coupling. We argue that although we expect the quantum critical point to get buried under a superconducting phase, signatures of the critical point may be visible in the finite temperature quantum critical regime. Our results have implications for experiments on proximity induced superconductivity in Dirac fermion systems, where they imply an enormous disorder-enhancement of the superconducting susceptibility. As a result, the proximity induced superconductivity in dirty systems is expected to be much stronger than that in clean systems at the Dirac point.
\end{abstract}
\maketitle

\section{Introduction}
The study of many body effects in disordered electronic systems has been a focus of condensed matter research for decades.\cite{Lee} An important subset of problems involves disordered electronic systems with attractive interactions, and hence the interplay of superconductivity and disorder.\cite{Abrikosov-Gorkov} The Anderson theorem\cite{Anderson} states that $s$-wave spin singlet superconductivity is immune to weak time reversal invariant disorder. However, disorder is widely believed to suppress all other forms of superconductivity. The Abrikosov-Gorkov theory\cite{Abrikosov-Gorkov} provides a framework for analyzing the suppression of superconductivity by disorder. More recently, it has been pointed out that mesoscopic fluctuation effects\cite{Spivak, Oreta, Lamacraft} can render superconductivity more robust to weak disorder than the Abrikosov-Gorkov theory would predict, but disorder nevertheless suppresses superconductivity.

The recent discovery of two-dimensional (2D) Dirac fermion systems, such as graphene\cite{Geim} or the surface of a 3D topological insulator,\cite{KaneHasan} has opened a new avenue of research into the interplay of disorder and superconductivity. The study of disordered Dirac fermions began almost 30 years ago \cite{FisherFradkin1985, Fradkin1986, Frad1986} and it is believed that a single species of Dirac fermions is protected against localization,\cite{KaneHasan} although a random scalar potential disorder generates a non-zero density of states.\cite{Frad1986, Ludwig}. Meanwhile, it is also known that Dirac fermion systems at charge neutrality do not develop superconductivity for arbitrarily weak attractive interactions, and that there is a quantum critical point at a critical attraction strength which separates the Dirac semimetal from the superconductor \cite{Wilson1973, GrossNeveu1974, Kopnin, Uchoa, Paramekanti, Juricic2013} In a recent stimulating development, it has been pointed out that this quantum critical point has an interesting effective field theory description, which displays emergent dynamical supersymmetry.\cite{grover, Sung-Sik} However, while disorder and attractive interactions have been studied in isolation for topological insulator surface states, the combination of disorder and attractive interactions has not been studied. 

In this paper, we study Dirac fermions with attractive interactions, in the presence of disorder. We show that the combination of scalar potential disorder and attractive interactions is particularly deadly to the Dirac semimetal, driving a transition into a superconducting phase for arbitrarily weak attraction strengths. Remarkably, for Dirac fermion systems disorder actually {\it enhances} superconductivity, allowing superconductivity to develop where the clean system would have been semi metallic. 
We also show that while the disorder enhancement of superconductivity can be estimated using a mean field theory \`{a} la Abrikosov-Gorkov, this treatment dramatically underestimates the strength of the superconductivity. A proper treatment of mesoscopic fluctuations is necessary to determine the strength of superconductivity in the disordered Dirac fermion system. Our results may also have implications for Dirac fermion systems with repulsive interactions and disorder \cite{Mirlin}. Dirac fermion systems with repulsive interactions are described by a Gross Neveu model, which also has a quantum critical point \cite{Gross} which may be destabilized by disorder. However, we do not pursue this line of research in this paper, leaving it as a topic for further work. We note too that whereas this paper is focused on Dirac semimetals, the interplay of disorder and superconductivity in normal metals  has been studied in Ref. \onlinecite{Feigelman}. 

While Dirac fermions arise both on the surface of a topological insulator and in graphene, in graphene the Dirac fermions are fourfold degenerate, coming in two spin and two valley flavors. On the surface of a topological insulator, however, there is just a single species of Dirac fermion. In this paper, we focus our analysis on the case of a topological insulator, with a single species of Dirac fermion. This captures the essential physics of Dirac fermions with attractive interactions and disorder, but is easier to treat analytically because of the smaller number of degrees of freedom. The basic conclusions should also apply to graphene, insofar as disorder will generate a non-vanishing density of states and enable superconductivity to develop for arbitrarily weak interactions. However, the existence of a valley degree of freedom in graphene, and the fact that disorder can cause intervalley scattering, may lead to additional features not present in the problem studied here. We leave the generalization of this analysis to graphene as a topic for future work. 

In this work, we consider a model of Dirac fermions with purely attractive, phonon mediated interactions, neglecting the Coulomb repulsion. In principle, the (unscreened) Coulomb interaction can prevent superconductivity \cite{Finkel2}, and the study of superconductivity in the presence of  Coulomb interactions is a subject we leave for future work. However, we note that in experiments on graphene or topological insulators, the Coulomb interaction can always be screened by metallic gates, so the neglect of the Coulomb repulsion can be justified. 

We also emphasize that our work has important implications for ongoing experiments attempting to induce superconductivity in graphene and topological insulators by means of the proximity effect. In the context of the proximity effect, the disorder enhancement of $T_c$ which we identify can be read as a disorder enhancement of the superconducting susceptibility. Thus, even in materials where there may not be an intrinsic attraction or intrinsic superconductivity, the proximity induced superconductivity in the disordered system will be dramatically stronger than than in the clean system, for reasons we explain in this paper.

 This paper is structured as follows. In Sec.~II, we discuss the zero temperature phase diagram of the clean Dirac fermion system in the presence of attractive delta function interactions. While the phase structure at the Dirac point (chemical potential $\mu=0$) as a function of attraction strength $g$ has previously been understood,\cite{Kopnin, Uchoa, Santos, Imada1, Imada2, Strack} the full phase diagram in the $g-\mu$ plane has not been presented as far as we are aware. Indeed, the $\mu \rightarrow \infty$ limit remains controversial, with the existing literature\cite{Santos, Imada1, Imada2} in apparent disagreement. We resolve this disagreement by means of a careful analysis that takes into account the finite ultraviolet cutoff for the interaction, which justifies use of a projected Hamiltonian. The projection operation introduces a gauge redundancy, which must be dealt with carefully. We also show how the system interpolates between the $\mu=0$ and large $\mu$ limits, and present the full zero temperature phase diagram. The zero temperature phase diagram contains two phases:  a Dirac semimetal along the $\mu=0$, $g<g_c$ line segment, and a superconductor everywhere else. There is a single superconducting phase, which is fully gapped and preserves time reversal symmetry everywhere. For small doping, the superconductivity is mostly spin singlet and $s$-wave. 
 
 In Sec.~III, we discuss the influence of disorder. First, we discuss disorder in the attraction strength $g$. We show that this form of disorder is a relevant perturbation at the $\mu=0, g=g_c$ critical point, and must necessarily change the universality class. However, we expect disorder in the attraction strength to be weak at the bare level, and thus the effect of attraction strength disorder may not manifest itself until very low energy scales. However, in any realistic experimental sample, there is likely to be significant disorder in the chemical potential $\mu$. We show that smooth chemical potential disorder destroys the semimetal phase, producing a network of electron and hole doped puddles. In the presence of attractive interactions, this system develops percolating superconductivity for arbitrarily weak attractive interactions, with a critical temperature that is exponentially small in the typical doping. We also show that short range disorder (which we model in terms of delta function impurities) also destroys the semimetal phase, introducing a low energy density of states that is exponentially small in the disorder strength. A uniform mean field analysis of the pairing problem suggests that the short range disordered Dirac fermion system should display superconductivity for arbitrarily weak interactions, but with a critical temperature that is doubly exponentially weak in the disorder. 
 
  In Sec.~IV, we analyze the influence of mesoscopic fluctuations on superconductivity for short range disorder, in the weak coupling limit. Our analysis suggests that mesoscopic fluctuations dramatically enhance superconductivity. In particular, the analysis suggests that small puddles of local superconductivity appear at temperatures that are linearly small in weak disorder, and the Josephson coupling between these puddles establishes global phase coherence at a temperature that is exponentially small in weak disorder. This represents a dramatic enhancement over the uniform mean field theory, which predicts a critical temperature that is doubly exponentially small in weak disorder. The application of a transverse magnetic field frustrates the global phase coherence, driving a phase transition into a gauge glass phase. 
 
 In Sec.~V, we analyze the interplay of disorder and attraction within a renormalization group (RG) framework, in the vicinity of the quantum critical point $\mu=0, g=g_c$. The construction of an RG for the attractive interaction requires us to work in an $\epsilon$-expansion about the upper critical spacetime dimension $D=4$. Meanwile, disorder is taken into account through a replica field theory approach. The discussion in this section complements the discussion in Sec.~III and IV. Whereas Sec.~III discussed the interplay of disorder and interactions near the free fermion point, the renormalization group analysis studies the interplay of disorder and interactions near the quantum critical point. We find that whereas chemical potential disorder is a marginally relevant perturbation at the Gaussian point, it is a (power law) irrelevant perturbation at the quantum critical point. Thus, one might naively expect the quantum critical point to be stable in the presence of weak chemical potential disorder. However, a careful analysis reveals that in fact chemical potential disorder is a dangerous irrelevant perturbation, in that it generates disorder in the BCS coupling, which is a {\it relevant} perturbation to the critical point. Thus, the supersymmetric critical point of Ref.~\onlinecite{grover, Sung-Sik} is in fact unstable in the presence of disorder. We expect that the flow away from the critical point leads to the superconducting phase, although firm conclusions cannot be drawn since the RG flows to strong coupling.

 Finally, in Sec.~VI we discuss the prospects of observing signatures of the supersymmetric quantum critical point identified in Ref.~\onlinecite{grover, Sung-Sik}, given the inevitability of disorder. We point out that even though the quantum critical point will be buried under the superconducting phase in any realistic experimental sample, in a sufficiently clean sample some signatures of the critical point may survive in the finite temperature quantum critical regime. We conclude by discussing future directions for the study of disorder and attractive interactions in Dirac fermion systems. 

We note that previous studies of Dirac quasiparticles in nodal superconductors have found a (secondary) superconducting transition in the disordered system, when no such transition occurs in the clean one \cite{Meyer, Gornyi, Florens, Dellanna}. There is some mathematical resemblance between these results and the discussion in Section III. However, the colossal enhancement of superconductivity by rare region effects discussed in Section IV, as well as the strong coupling physics discussed in Section V,VI, have no analog in these works. The discussion in section III also has important differences, in that it belongs to a different Cartan symmetry class \cite{Schnyder}, with very different localization physics, and also in that we are discussing a true superconducting instability in a semimetal, whereas the works \cite{Meyer, Gornyi, Florens, Dellanna} are discussing the appearance of a secondary s-wave component of the order parameter in a d-wave superconductor.

\section{Phase diagram of clean Dirac fermions with attractive interactions}

We begin by considering a single species of Dirac fermions in the absence of disorder, with short range attractive interactions. 
The creation operator for Dirac fermions with momentum $\hbar \vec{k}$ is $\psi^{\dag}_{\vec{k}}$. The spinor structure of the Dirac fermions is implemented by defining $\psi^{\dag}_{\vec{k}} = (c^{\dag}_{\uparrow,\vec{k}}, c^{\dag}_{\downarrow,\vec{k}})$, where $c^{\dag}_{\sigma, \vec{k}}$ creates a fermion with spin $\sigma$ and wavevector $\vec{k}$. It is convenient to introduce the Pauli matrices $\sigma_i$ which act in the spin space, and to also define $\sigma_0$ to be the two dimensional identity matrix acting in spin space. The second quantized Hamiltonian may then be written as $H = H_0 + H_1$, where 
%
\begin{eqnarray}
H_0 &=& \sum_{\vec{k}} \psi^{\dag}_{\vec{k}} \left(-\mu\sigma_0 + v k_x \sigma_1 + v k_y \sigma_2 \right) \psi_{\vec{k}}, \label{eq: H0} \\
H_1 &=& \sum_{\vec{k,p,q}} V(\vec{q}, \vec{k}, \vec{p}) \psi^{\dag}_{\vec{k}} \sigma_0 \psi_{\vec{k+q}} \psi^{\dag}_{\vec{p}} \sigma_0 \psi_{\vec{p-q}}, \label{eq: H1}
\end{eqnarray}
where $v$ is the Fermi velocity, $\mu$ is the chemical potential (which controls the doping level), and we have assumed a purely `density-density' interaction, which has no intrinsic spin structure. The interaction is assumed to be attractive, $V<0$. Although we have taken $\sigma$ to be a spin index for simplicity, we are aware that for generic topological insulator surfaces it may be a composite spin-sublattice index \cite{ZKM}. The distinction makes no difference to our analysis, which involves pairing between time reversed states (not necessarily opposite spin states). The Pauli matrices should thus be understood as acting in the spin/pseudospin space relevant for the surface states, such that $\sigma$ and $-\sigma$ are time reversed states.

We now project onto the Cooper channel, by restricting the interaction Hamiltonian (\ref{eq: H1}) to $\vec{p} = -\vec{k}$. The projection is appropriate for studying the superconductivity of Dirac fermions. The projection leads to a Hamiltonian of the form $H_0$ + $H_1'$, where $H_0$ is given by (\ref{eq: H0}) and 
\begin{equation}
H_1' =  \sum_{\vec{k,q}} V(\vec{q},\vec{k}) \psi^{\dag}_{\vec{k}} \sigma_0 \psi_{\vec{q}} \psi^{\dag}_{\vec{-k}} \sigma_0 \psi_{\vec{-q}}.\label{eq: H1'}
\end{equation}
We further restrict ourselves to a short range, $\delta$ function interaction. In this case, $V$ is independent of momenta and we obtain the BCS Hamiltonian for Dirac fermions
\begin{eqnarray}
H_\textrm{BCS} &=& \sum_{\vec{k}} \psi^{\dag}_{\vec{k}} \bigg(-\mu\sigma_0 + v k_x \sigma_1 + v k_y \sigma_2 \bigg) \psi_{\vec{k}}\nonumber\\ &-&   g\sum_{\vec{k,q}} \psi^{\dag}_{\vec{k}} \sigma_0 \psi_{\vec{q}} \psi^{\dag}_{\vec{-k}} \sigma_0 \psi_{\vec{-q}}, \label{eq: HBCS}
\end{eqnarray}
where $g = -V$ is the superconducting coupling. The rest of this section will be devoted to solving (\ref{eq: HBCS}). 

We proceed as follows. First, we provide a general discussion aimed at classifying potential solutions of (\ref{eq: HBCS}) into distinct phases. Then we solve (\ref{eq: HBCS}) along the line $\mu=0$. The solution along this line is well known, but we present it for completeness sake. Next, we discuss superconductivity in the limit $|\mu| \rightarrow \infty$. Our discussion resolves a disagreement between Ref.~\onlinecite{Santos} and Ref.~\onlinecite{Imada1, Imada2} as to the nature of superconductivity in this limit, and highlights subtleties connected with UV cutoffs and gauge ambiguities which must be properly taken into account to understand this limit. Finally, we solve (\ref{eq: HBCS}) for arbitrary doping, and construct the full phase diagram in $\mu-g$ space. 

We show that the entire $\mu-g$ plane has a ground state that belongs to a single superconducting phase, with the exception of a single line segment along the $\mu=0$ line, which is semi-metallic. This follows because the Hamiltonian (\ref{eq: HBCS}) has a superconducting ground state for any system with a non-vanishing density of states, for arbitrarily weak interactions, and the low energy density of states vanishes only on the $\mu=0$ line. More non-trivial is the fact that the $\mu>0$ and $\mu<0$ superconducting regions belong in the same phase - a phase which is fully gapped and time reversal invariant, with an order parameter that is a real linear combination of spin singlet and spin triplet parts. A spin triplet component emerges because the spin singlet order induces spin triplet order at any non-zero doping $\mu \neq 0$. 

\subsection{Symmetries and superconductivity}
\label{sec: phases}
The most general order parameter contains spin singlet and triplet pieces, and can be written as 
\[
\Delta = \sum_{\vec{k}} \langle \psi_{\vec{k}}(\Delta_{s,\vec{k}} + \vec{d}_{\vec{k}} \cdot \vec{\sigma}) i\sigma_2 \psi_{\vec{-k}} \rangle.
\]
Fermi statistics demand that $\Delta_{s,\vec{k}}=\Delta_{s,-\vec{k}}$ and $d_{\vec{k}} = -d_{-\vec{k}}$. The Hamiltonian is invariant under time reversal (TRS), which is implemented by complex conjugation, taking $\vec{k}\rightarrow -\vec{k}$, and acting with the operator $i\sigma_2$. Thus, we conclude that TRS is preserved if and only if $\vec{d_k}/\Delta_s$ is a vector with purely real components (i.e. if the superconductivity is a real linear combination of singlet and triplet). There is also a particle hole symmetry along the $\mu=0$ line and we will discuss this symmetry when appropriate. However, this symmetry is broken by scalar potential disorder (our primary focus in this paper), and so we do not attach much weight to this symmetry. However, there is a second meaningful distinction between superconducting phases,\cite{Santos, Nagaosa} namely whether there exist gapless Bogoliubov-de Gennes (BdG) quasiparticles. As we shall see, the superconducting solutions to (\ref{eq: HBCS}) are always fully gapped and invariant under time reversal, so the phase diagram contains a single superconducting phase. 

We assume that spin singlet superconductivity is the primary superconducting instability. We now discuss whether spin singlet order can induce spin triplet order. 
At non-zero $\mu$, the only symmetries are TRS (discussed above), and also a continuous rotation symmetry implemented by the generator $J_z = L_z + \frac12 \sigma_3$. Spin singlet order is even under TRS, and has $J_z = 0$. Time reversal symmetry prohibits the spin singlet order from coupling to any spin triplet order parameter with complex $\vec{d_k}/\Delta_s$, whereas rotation symmetry prohibits it from coupling to any spin triplet order with $J_z \neq 0$. However, a spin triplet order parameter with $\vec{d_k}/\Delta_s = \vec{k}$ is even under TRS, has $J_z = 0$, and can thus couple directly to the spin singlet order parameter. Thus, in general we expect a solution of the form 
\begin{equation}
\Delta = \sum_{\vec{k}} \langle \psi_{\vec{k}}(\Delta_{s} + F(\mu) \vec{k} \cdot \vec{\sigma}) i\sigma_2 \psi_{\vec{-k}} \rangle, \label{eq: general solution}
\end{equation}
where $\vec{k} = (k_x, k_y)$ and the spin triplet piece of the order parameter is induced by the spin singlet piece. The proportionality constant $F(\mu)$ remains to be determined. 

The Dirac point $\mu = 0$ is special, in that a large number of extra symmetries appear. In particular, the theory becomes Lorentz invariant, and also there exists at $\mu = 0$ a particle hole symmetry,\cite{Schnyder} under which spin singlet and spin triplet pairing are even and odd respectively. Thus, the spin singlet and spin triplet orders cannot couple at $\mu = 0$, and it follows that $F(\mu=0) = 0$. 

The remainder of this section will be devoted to explicitly obtaining and solving BdG equations leading to Eq.(\ref{eq: general solution}). We will show that $F(\mu \rightarrow \pm \infty) = \sign(\mu)$ and $F(\mu \rightarrow 0) \sim g' \mu$, where $g'$ is the ($\mu$ dependent) attraction in the triplet channel. Readers uninterested in the clean system details may skip directly to Sec.~III.

\subsection{Superconductivity along the $\mu = 0$ line}
We wish to solve $H_\textrm{BCS}$ at $\mu=0$. Since the Pauli principle ensures that two fermions with the same spin cannot interact through a delta-function potential, we can re-write the Hamiltonian as 
\begin{eqnarray}
H &=& \sum_{\vec{k}} \psi^{\dag}_{\vec{k}} \bigg(v k_x \sigma_1 + v k_y \sigma_2 \bigg) \psi_{\vec{k}}\nonumber\\ &-&   g\sum_{\vec{k,q}} \psi^{\dag}_{\vec{k}} (-i \sigma_2) \psi^{\dag}_{-\vec{k}} \psi_{\vec{q}}  i\sigma_2 \psi_{\vec{-q}}. \label{eq: mu0}
\end{eqnarray}
Superconductivity is necessarily spin singlet, and is characterized by the order parameter 
\begin{equation}
\Delta_{s} = \sum_{\vec{k}} \langle \psi_{\vec{k}} (i\sigma_2) \psi_{-\vec{k}} \rangle.\label{eq: deltas}
\end{equation}
It is now convenient to introduce the Euclidean time path integral representation of the partition function, $Z = \int D[\psi^{\dag}, \psi] \exp\big(- \int_0^{\beta} d\tau \int d^2x L[\psi^{\dag}, \psi]\big)$, where $\beta$ is the inverse temperature and the Lagrangian takes the form 
\[
L = \int \frac{d^2k}{(2\pi)^2} \psi^{\dag}_{\vec{k}} \partial_{\tau} \psi_{\vec{k}} + H,
\]
where we have replaced the sum over momenta by an integral. Since we are working in the path integral representation, $\psi^{\dag}$ and $\psi$ now represent Grassman valued fields rather than second quantized operators. 
We define the four component Nambu spinor $\Psi = (\psi_{\vec{k}}, \psi^{\dag}_{-\vec{k}})$. After decoupling the four-fermion interaction in (\ref{eq: mu0}) by means of a Hubbard-Stratonovich transformation, the partition function can be rewritten as $Z = \int D[\Psi^{\dag}, \Psi, \Delta^*, \Delta] \exp\big(- \int_0^{\beta} d\tau \int d^2x L[\Psi^{\dag}, \Psi, \Delta^*, \Delta]\big)$, where $\Delta$ is a complex valued (bosonic) field, and 
\begin{align}
&L = T \sum_{\omega_n} \int \frac{d^2k}{(2\pi)^2} \Psi^{\dag}_{\omega_n, \vec{k}} G_{\omega_n, \vec{k}} \Psi_{\omega_n, \vec{k}} + \frac{|\Delta_{s}|^2}{2 g},\nonumber \\
&G_{\omega_n, \vec{k}} = \left(\begin{array}{cccc} i\omega_n  -\mu & v k_+ &0 & \Delta_s \\ v k_- & i\omega_n - \mu & -\Delta_s &0 \\ 0& -\Delta_s^* & i\omega_n + \mu & v k_- \\ \Delta_s^* & 0& v k_+ & i\omega_n + \mu \end{array} \right),\nonumber
\end{align}
where we have introduced the Matsubara frequencies $\omega_n = (2n+1)\pi T$, and have assumed that the $s$-wave order parameter $\Delta_s$ is isotropic in momentum space. 

We can now integrate out fermions exactly to obtain a Lagrangian that only involves the order parameter fields. This Lagrangian takes the form
\begin{eqnarray}
L &=& - T \sum_{\omega_n} \int \frac{d^2k}{(2\pi)^2}\Tr \ln G_{\omega_n,\vec{k}} + \frac{|\Delta_{s}|^2}{2 g}\nonumber\\
&=& - T \sum_{\omega_n} \int \frac{d^2k}{(2\pi)^2} \ln(\omega^2 + v^2 k^2 + |\Delta_s|^2)^2 
+ \frac{|\Delta_{s}|^2}{2 g}. \nonumber
\end{eqnarray}
Variation with respect to $|\Delta|^2$ then yields the gap equation,
\[
\frac{2 T}{2\pi} \sum_n \int \frac{k dk}{\omega_n^2 + v^2 k^2 + |\Delta_s|^2} = \frac{1}{2g}.
\]
The momentum integrals carry an implicit cutoff at the scale $v k = \omega_D$, where $\omega_D$ is the Debye frequency. At zero temperature $T\sum_{\omega_n} \rightarrow \int \frac{d \omega}{2\pi}$, and the gap equation can be solved as 
\[
\sqrt{\omega_D^2+|\Delta_s|^2} - |\Delta_s| = 2\pi v^2/g,
\]
which has solutions only for 
\[
g > g_c = 2\pi v^2/\omega_D.
\]
Thus we recover the well known result that superconductivity for Dirac fermions along the $\mu=0$  line is a threshold phenomenon, with the order parameter developing a non-zero expectation value only if interactions are strong enough. The solutions (assuming $|\Delta_s| \ll \omega_D$) take the form
\[
\qquad |\Delta_s| = \omega_d - \frac{2\pi v^2}{g} \approx \frac{2\pi v^2}{g_c^2} \delta g,
\]
where the approximate equality holds close to the threshold, $\delta g = g-g_c \ll g_c$, but not so close as to be governed by the critical point. Thus, we conclude that along the $\mu=0$ line, the system has spin singlet superconducting order for $g>g_c$ and is a semimetal for $g<g_c$. As has been pointed out in Ref.~\onlinecite{grover, Sung-Sik}, the critical point $g=g_c$ has some unusual features, and is described by an effective field theory that exhibits  emergent dynamical supersymmetry.

\subsection{Superconductivity at large doping, $\mu \gg \omega_D$}

At large doping, the spin basis is not the most convenient basis to work with, since the low energy states are linear superpositions of spin up and spin down states. Instead, we transform to a basis of $+$ and $-$ helicity states (upper and lower Dirac cones) by performing a unitary transformation, according to
\begin{align}
\Phi_{\vec{k}} &= \left( \begin{array}{c} c_{+,\vec{k}} \\ c_{-,\vec{k}} \end{array}\right) = U \psi_{\vec{k}}, \nonumber\\
 U &=  \frac{e^{i G \phi_\vec{k}}}{\sqrt{2}}\left( \begin{array}{cc} e^{i\phi_\vec{k}/2} & e^{-i\phi_\vec{k}/2} \\ e^{i\phi_{\vec{k}}/2} & - e^{-i \phi_{\vec{k}}/2} \end{array}\right), \label{eq: unitary}
\end{align}
where $k_x + i k_y = |\vec{k}| e^{i\phi_\vec{k}}$. The requirement that $U$ should be single valued demands that $G$ should be a half integer. We wish to stress that the choice of unitary matrix in (\ref{eq: unitary}) is not unique, since we can freely choose $G$ to be any half integer. This `gauge ambiguity' makes no difference if we work with the full Hamiltonian. However, we will shortly be projecting onto a single helicity basis, and the projected Hamiltonian will look different with different gauge choices. The different projected Hamiltonians should be understood as being gauge equivalent. 

We now use the unitary transformation (\ref{eq: unitary}) to express the Hamiltonian in the helicity basis. Defining $\tau_0$ to be the identity matrix in helicity space, and $\tau_i$ to be the Pauli matrices in helicity space, we can rewrite the Hamiltonian in the helicity basis as 
\begin{align}
&H = \sum_{\vec{k}} \Phi^{\dag}_{\vec{k}} (-\mu\tau_0 + v k \tau_3) \Phi_{\vec{k}} - g \sum_{\vec{k,q}} \frac{e^{2 i G (\phi_\vec{q} - \phi_\vec{k})}}{4} \nonumber\\ &\times\Phi^{\dag}_{\vec{k}} \left[2 \cos\left(\frac{\phi_\vec{k}-\phi_\vec{q}}{2} \right) \tau_0 + 2 i \sin \left( \frac{\phi_\vec{k} -\phi_\vec{q}}{2} \right) \tau_1 \right]\Phi_{\vec{q}}\nonumber\\ 
&\times \Phi^{\dag}_{\vec{-k}} \left[2 \cos\left(\frac{\phi_\vec{k}-\phi_\vec{q}}{2} \right) \tau_0 + 2 i \sin \left( \frac{\phi_\vec{k} -\phi_\vec{q}}{2}\right) \tau_1 \right]\Phi_{\vec{-q}},\nonumber \\ \label{eq: Hhelicity}
\end{align}
where we have not yet specified the choice of gauge $G$. 

It is intuitively obvious that at large doping, only states close to the Fermi surface need to be considered, and thus one can project onto the states with helicity $\sign( \mu)$. To justify this projection, we note that the interaction implicitly has an ultraviolet cutoff on the scale $\omega_D$. Thus, it cannot couple states near the Fermi surface to states further than $\omega_D$ away from the Fermi surface. If the doping $|\mu|>\omega_D$, then one can project onto states with a single helicity. The apparent neglect of this UV cutoff in Ref.~\onlinecite{Santos} explains the discrepancy between the large doping results in Ref.~\onlinecite{Santos} and Ref.~\onlinecite{Imada1,Imada2}. 

We consider electron doping $\mu>0$. The case of hole doping $\mu<0$ follows by analogy. After projection onto states with positive helicity, the Hamiltonian becomes
\begin{align}
H &= \sum_{\vec{k}} c^{\dag}_{+,\vec{k}} \big(-\mu+ v k \big) c_{+,\vec{k}}\nonumber\\ 
&-g  \sum_{\vec{k,q}}e^{2iG(\phi_\vec{q}-\phi_\vec{k})} \cos^2\left(\frac{\phi_\vec{k} - \phi_\vec{q}}{2} \right)\nonumber\\
&\times c^{\dag}_{+,\vec{k}} c_{+,\vec{q}} c^{\dag}_{+,\vec{-k}} c_{+,-\vec{q}}.\nonumber
\end{align}
Now $\cos^2{x/2} = \frac12(1 + \cos(x)) = \frac14(2+ e^{ix}+e^{-ix})$. Thus, note that the projected interaction has harmonics with angular momentum $2G, 2G+1, 2G-1$, where $G$ can be any half integer. It is most convenient to make the gauge choice $G=1/2$. Then the attractive potential has harmonics with angular momenta $l =0,1,2$. Note that an effective `$p$-wave' (l=1) harmonic has been generated by the projection, even though we started with a purely $s$-wave interaction.\cite{Fu+Kane}

Now since after projection we are dealing with a standard one band BCS problem for spinless fermions, Fermi statistics demand that the superconductivity has to be odd parity i.e. $\Delta_{\vec{-k}} = - \Delta_{\vec{k}}$. Thus, the $l=0,2$ harmonics do not introduce superconductivity, and may be projected out. We need retain only the $p=1$ harmonic, which gives us a Hamiltonian\cite{Fu+Kane}
\begin{eqnarray}
H &=& \sum_{\vec{k}} c^{\dag}_{+,\vec{k}} \big(-\mu+ v k \big) c_{+,\vec{k}}\nonumber\\ &-&g  \sum_{\vec{k,q}}\frac{e^{i(\phi_\vec{q}-\phi_\vec{k})}}{2} c^{\dag}_{+,\vec{k}} c_{+,\vec{q}} c^{\dag}_{+,\vec{-k}} c_{+,-\vec{q}}. \nonumber
\end{eqnarray}
%
Proceeding to the path integral representation, and decoupling the four fermion interaction using an order parameter field $\Delta_{+} = \frac{g}{2} \sum_{\vec{q}} \langle c_\vec{q} c_{-\vec{q}}\rangle e^{ i \phi_\vec{q}}$, we obtain the Euclidean time Lagrangian

\begin{eqnarray}
L &=& \sum_{\vec{k}} c^{\dag}_{+,\vec{k}} \big(\partial_{\tau}-\mu+ v k \big) c_{+,\vec{k}} \nonumber\\ &+& \sum_{\vec{q}} (\Delta_{+} c_\vec{q} c_{-\vec{q}} e^{i\phi_\vec{q}}  + \textrm{c.c.}) + \frac{1}{g} |\Delta_{+}|^2.\nonumber
\end{eqnarray}
After integrating out the fermions, we obtain an action purely in terms of order parameter fields, which takes the form
\[
L = - \Tr \ln \bigg[ \omega_n^2 + (vk- \mu)^2 + |\Delta_{+}|^2 \bigg] + \frac{1}{g} |\Delta_{+}|^2,
\]
where $\Tr$ denotes summation over Matsubara frequencies and integration over momenta. Variation with respect to $|\Delta_{+}|^2$ yields the gap equation,
\[
\Tr \frac{1}{\omega_n^2 + (vk- \mu)^2 + |\Delta_{+}|^2} = \frac{1}{g},
\]
and summing over fermonic Matsubara frequencies gives
\[
\int \frac{d^2k}{(2\pi)^2} \frac{1}{2 \sqrt{(vk-\mu)^2 + |\Delta_+|^2}} \tanh \left(\frac{\xi_\vec{k}}{2T}\right)=\frac{1}{g},
\]
where the integral goes over $\mu-\omega_D < vk < \mu + \omega_D$, and we are working in the limit $\mu \gg \omega_D$. In this limit, the zero temperature gap equation can be solved as 
\[
\ln \bigg( \frac{\omega_D + \sqrt{\omega_D^2 + |\Delta_+|^2}}{\Delta_+}\bigg) = \frac{2\pi v^2}{g \mu}.
\]
In the weak coupling limit $2\pi v^2/g\mu \gg 1$, this gives us 
\begin{equation}
|\Delta_+| = 2 \omega_D \exp\left(-\frac{2\pi v^2}{g\mu} \right). \label{eq: largemu}
\end{equation}
Thus one concludes that there is an order parameter $\Delta_{+}(\vec{k}) = \frac{g}{2} \sum_{\vec{q}} \langle c_\vec{q} c_{-\vec{q}}\rangle e^{ i \phi_\vec{q}}$ with expectation value given by (\ref{eq: largemu}). For a different gauge choice, the expression (\ref{eq: largemu}) still holds, but the order parameter instead has the form $\Delta_{+}(\vec{k}) = \frac{g}{2} \sum_{\vec{q}} \langle c_\vec{q} c_{-\vec{q}}\rangle e^{ i2G \phi_\vec{q}}$.  The calculation for $\mu<0$ is analogous, with the only difference being that the projection is now on the band with negative helicity. Thus, we conclude that in the helicity basis, the order parameter takes the form 
\[
\Delta = \sum_\vec{k} e^{ 2i G\phi_\vec{k}} \langle \Phi^T_\vec{k} (\tau_0 + \sign(\mu) \tau_3) \Phi_{-\vec{k}}\rangle.
\]
The BdG spectrum is fully gapped. Meanwhile, transforming back to the spin basis, the order parameter becomes 
\begin{equation}
\Delta = \sum_\vec{k} \langle \Psi^T_\vec{k} \big(1 +\sign(\mu) ( \cos \phi_\vec{k} \sigma_1 - \sin \sigma_2)\big)i\sigma_2\Psi_{-\vec{k}}\rangle. \label{eq: OPbigmu}
\end{equation}
Note that this is a real linear combination of spin singlet and spin triplet terms, and thus preserves time reversal symmetry. Thus, the superconducting state at large doping is in the same phase (fully gapped, time reversal invariant) as the superconducting phase at $\mu=0$. It seems likely that the system should smoothly interpolate between the $\mu=0$ and large $\mu$ limits. We confirm this in the next section. 

\subsection{Superconductivity at small, nonzero $\mu$}
We have solved the Hamiltonian (\ref{eq: HBCS}) in the $\mu=0$ and large $\mu$ limits. We now turn to the small $\mu$ regime. We have seen that a finite UV cutoff $\omega_D$ on the interaction generates a spin triplet component to the interaction away from $\mu=0$. We therefore introduce a weak spin triplet component into the action by hand, and write the Lagrangian (after Hubbard-Stratanovich decomposition) as 
%
%
\begin{widetext}
 \begin{equation}
L = \sum_{\vec{k}} \Psi^{\dag}_{\vec{k}}  \left(\begin{array}{cccc} i\omega_n  -\mu & v k_+ & -d_1(\vec{k})+i d_2(\vec{k}) & \Delta_s  +  d_3(\vec{k})\\ v k_- & i\omega_n - \mu & -\Delta_s  +  d_3(\vec{k}) & d_1(\vec{k}) + i d_2(\vec{k}) \\ -d_1^*(\vec{k}) - i d_2^*(\vec{k}) & -\Delta_s^*+ d_3^*(\vec{k}) & i\omega_n + \mu & v k_- \\ \Delta_s^*+ d^*_3(\vec{k}) & d^*_1(\vec{k}) - i d^*_2(\vec{k}) & v k_+ & i\omega_n + \mu \end{array} \right)\Psi_{\vec{k}} + \frac{|\Delta_{s}|^2}{2 g} + \frac{\sum_{i=1}^3 |d_{i}|^2}{2 g'},\nonumber
\end{equation}
\end{widetext}
where $g'$ is a function of $|\mu|$. Consistency with the $\mu=0$ and large $\mu$ limits requires that $g'(0)=0$ and $g'(|\mu|>  \omega_D) = g$. We assume that $g'$ interpolates smoothly in between, so that $0<g'<g$ in the small $\mu$ regime. Meanwhile, the Matsubara frequencies are implicitly summed over,  and $d_{1,2,3}$ are odd functions of $\vec{k}$. 

After integrating out fermions, we obtain
\begin{widetext}
\begin{eqnarray}
L &=& - \Tr \ln \Bigg( (\omega^2 + k^2)^2 + 2 (\omega^2 + k^2) |\Delta_{s}|^2  + 2(\omega^2 - k^2) |d_3(\vec{k})|^2 + 2 (\omega^2 - k_x^2 + k_y^2)|d_2(\vec{k})|^2 + 2 (\omega^2 + k_x^2 - k_y^2) |d_1(\vec{k})|^2\nonumber\\&& - 4 k_x k_y (d_1(\vec{k}) d_2^*(\vec{k}) + d_1^*(\vec{k}) d_2(\vec{k})) + 4 w k_x ( d_3(\vec{k}) d_2^*(\vec{k}) + \textrm{c.c.}) + 4 w k_y (d_3(\vec{k}) d_1^*(\vec{k}) + \textrm{c.c.})+  |\Delta_s|^4 + \sum_{i=1}^3 |d_i(\vec{k})|^4  \nonumber\\&&- \bigg[\Delta_s^2  \big[ (d_1^*(\vec{k}))^2 +  (d^*_2(\vec{k}))^2 + (d_3^*(\vec{k}))^2 \big]+ \textrm{c.c.}\bigg] + \bigg[d_1(\vec{k})^2 d^*_3(\vec{k})^2  +  d_1(\vec{k})^2 d^*_2(\vec{k})^2 +  d_3(\vec{k})^2 d^*_2(\vec{k})^2  + \textrm{c.c.}\bigg] \nonumber \\&&- 4 k_y \mu \big(\Delta_s^*  d_2(\vec{k}) + \Delta_s  d_2^*(\vec{k})\big) + 4 k_x \mu \big( \Delta_s^* d_1(\vec{k}) + \Delta_s d_1^*(\vec{k})\big) + 2 \mu^2 \big(|\Delta_s|^2  + \sum_i d_i^2 + \omega^2 - k^2\big) +  \mu^4\Bigg)\nonumber\\
&&+ \frac{|\Delta_{s}|^2}{2 g} + \frac{\sum_{i=1}^3 |d_{i}|^2}{2 g'},\label{eq: bigLagrangian}
\end{eqnarray}
\end{widetext}
where we have set $v=1$ for simplicity. Now, $g>g'$ in this regime, so that $\Delta_s$ is the primary instability. Thus, we first set $\vec{d}=0$ and solve for $\Delta_s$, and then use this value of $\Delta_s$ to solve for $\vec{d}$. Setting $\vec{d}=0$ in the above action, varying with respect to $|\Delta_s|^2$, and linearizing in small $|\Delta_s|$ by working near $T_c$, we obtain the linearized gap equation,
\[
\frac{T_c}{\pi} \sum_n \int \frac{k dk (\omega_n^2 + k^2 + \mu^2)}{(\omega_n^2 + k^2 + \mu^2)^2 -4 \mu^2 k^2} = \frac{1}{2g},
\]
where $\omega_n$ are fermonic Matsubara frequencies $\omega_n = (2n+1)\pi T_c$. In the strong coupling limit $g\rightarrow \infty$ the presence of a non-zero $\mu$ is inessential and we can just use the strong coupling solution at $\mu=0$. However, in the weak coupling limit $g\rightarrow 0$ the chemical potential is vitally important. At $\mu = 0$, there is no superconductivity, but as we shall discover, superconductivity develops for any non-zero $\mu$. To see this, we sum over Matsubara frequencies to obtain the new gap equation,
%
\[
\int_0^{\omega_D}  k d k \bigg[\frac{\tanh \big(|k+ \mu|/2T_c\big)}{|k+\mu|} + \frac{\tanh \big(| k- \mu|/2T_c \big)}{| k- \mu|} \bigg] = \frac{\pi}{g},
\]
which defines $T_c$. Assuming $\mu \ll \omega_D$, and working in the weak coupling regime $g \mu \ll 1$, this has solution
\begin{equation}
T_c \sim \omega_D\exp\left( - \frac{\pi}{g|\mu|} \right). \label{eq: Tcsmallmu}
\end{equation}
Thus, we find there is a weak coupling instability to spin singlet superconductivity at non-zero temperature for any non-zero $\mu$. However, the dependence on $\mu$ is strongly non-analytic. The expression (\ref{eq: Tcsmallmu}) can be understood by noting that it is just the solution of the standard BCS Hamiltonian for a system with a low energy density of states proportional to $\mu$.  The zero temperature gap $|\Delta_s(T=0)|$ is proportional to $T_c$. 

Thus, we have determined the magnitude of the primary superconducting order parameter $\Delta_s$. We now determine what happens to the triplet fields $\vec{d}$ in the presence of the non-zero $|\Delta_s|$. To see this, we expand the Lagrangian (\ref{eq: bigLagrangian}) in small $\vec{d}$. Close to $T_c$ (\ref{eq: Tcsmallmu}), the expansion takes the form
\begin{eqnarray}
L &=& L_0 + \sum _{ij}\alpha_{ij} d_i^* d_j + \Tr F 4 k_y \mu \big(\Delta_s^* d_2(\vec{k}) + \Delta_s  d_2^*(\vec{k})\big)\nonumber \\&-& \Tr F 4 k_x \mu \big( \Delta_s^* d_1(\vec{k}) + \Delta_s d_1^*(\vec{k})\big),\nonumber
 \end{eqnarray}
where $L_0$ is the Lagrangian at $\vec{d}=0$, the matrix $\alpha_{ij}$ has positive definite eigenvalues $\sim 1/g'$, and $F$ is a strictly positive function of frequencies and momenta, which is even in $\omega$ and $k$. The positivity of the eigenvalues of $\alpha_{ij}$ ensures that the critical temperature for spin triplet order to develop in the absence of spin singlet order is lower than the $T_c$ (\ref{eq: Tcsmallmu}). This condition must be satisfied in order for $\Delta_s$ to be the primary instability. 

Now note that at finite $\mu$, a non-zero $\Delta_s$ automatically generates a nonzero $d_{1,2} \sim \mu \Delta_s$. This can be straightforwardly verified by minimizing with respect to $d_i$. In effect, $\Delta_s$ acts as a symmetry breaking field for the $d_i$ at non-zero chemical potential. Moreover, the structure of the Landau expansion picks out $d_1 \sim k_x/k =  \cos \theta_\vec{k}$ and $d_2 \sim k_y/k = \sin \theta_\vec{k}$, so that the integrals when multiplied by $k_x$ and $k_y$ respectively are non vanishing. Thus, we find that at small $\mu$, the superconducting order parameter takes the form
\begin{equation}
\Delta_{+} = \sum_\vec{k} \langle \Psi^T_\vec{k} \big(1 +K g' \mu ( \cos \phi_\vec{k} \sigma_1 - \sin \sigma_2)\big)i\sigma_2\Psi_{-\vec{k}}\rangle, \label{eq: OPsmallmu}
\end{equation}
where $K$ is some undetermined positive constant and where $g'$ is a function of $g, \mu$ and $\omega_D$. The attraction in the triplet channel, $g'$, may already be non-zero at the bare level, and it can be verified that a $g'$ interaction is also generated at second order in perturbation theory in small $g$. If $g'(\mu=0)=0$, then the leading contribution to $g'$ appears to scale as $g' \sim \mu^2 g^2 / \omega_D^3$. However, we have not verified this result and a rigorous determination of $g'(g, \mu,\omega_D)$ lies beyond the scope of this work. Moreover, the behavior of $g'(g, \mu, \omega_D)$ at strong coupling and near the critical point $g=g_c$ lies beyond the reach of perturbation theory in $g$, and may be a interesting topic for further work. Near the critical point $g=g_c$, the expression (\ref{eq: OPsmallmu}), which is based on mean field theory, may also be invalid, and the nature of the singlet-triplet coupling near the critical point is an interesting topic for future study. For our present purposes, it is sufficient to note that we can smoothly go from the time reversal symmetric, gapped superconducting state at $\mu=0$ to the time reversal symmetric, gapped state at large $\mu$, according to (\ref{eq: OPsmallmu}).  


The phase diagram follows straightforwardly. The line segment $\mu=0, g<g_c$ is semi metallic, and everywhere else in the $\mu-g$ plane there is a single superconducting phase, which is gapped and time reversal symmetric. The structure of the order parameter evolves smoothly according to 
\begin{equation}
\Delta_{+} = \sum_\vec{k} \langle \Psi^T_\vec{k} \big(1 +F(\mu) ( \cos \phi_\vec{k} \sigma_1 - \sin \sigma_2)\big)i\sigma_2\Psi_{-\vec{k}}\rangle, \label{eq: OPallmu}
\end{equation}
where $F(\mu)$ interpolates smoothly between the limits $F(\mu) = K g'(g,\mu,\omega_D) \mu$ for small $\mu$ and $F(\mu) = \sign(\mu)$ for large $|\mu|>\omega_D$.

\begin{figure}
\includegraphics[width = \columnwidth]{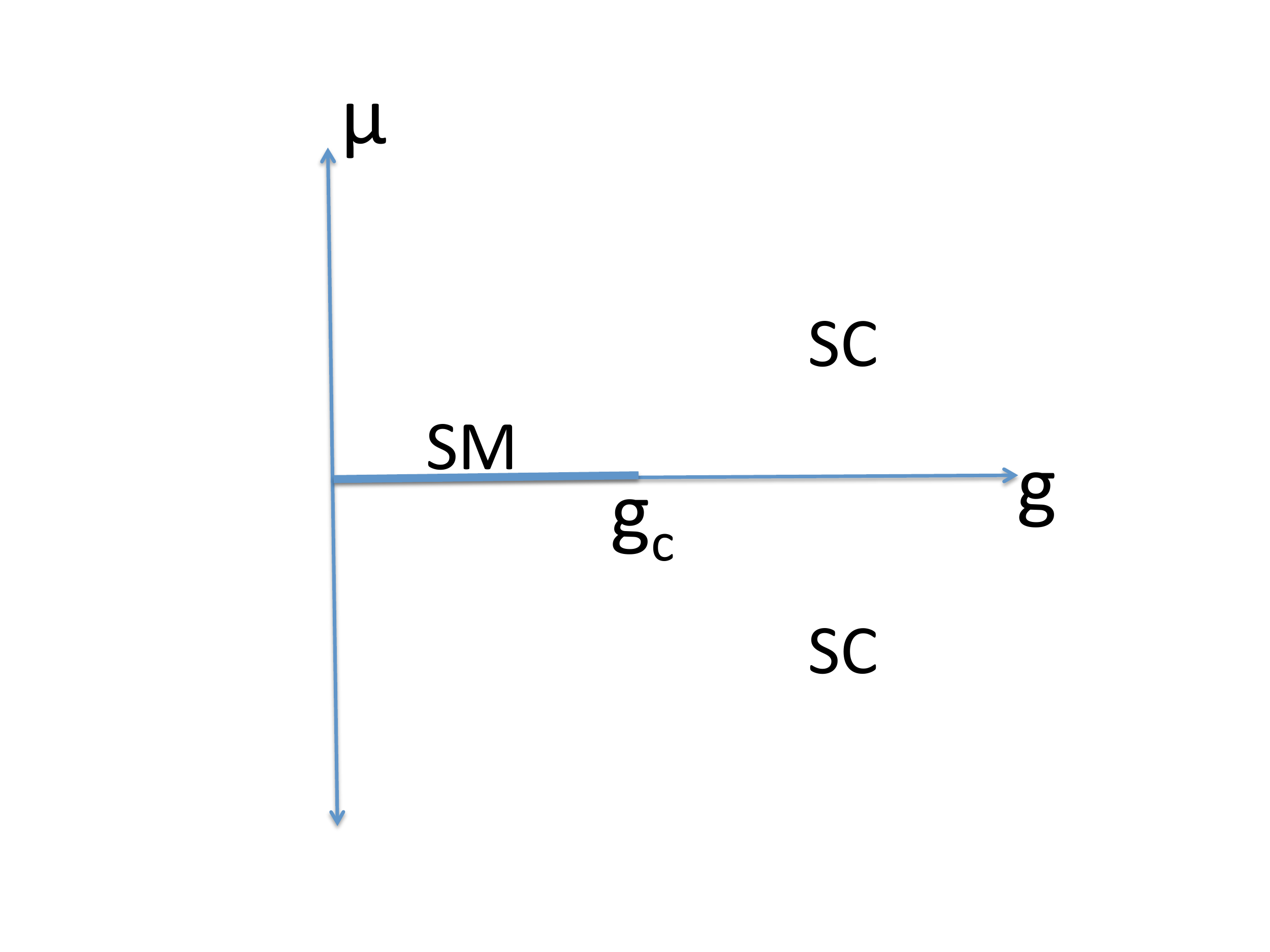}
\caption{Phase diagram of clean Dirac fermion system. Along the line segment $\mu=0$, $g<g_c$ the semimetal phase (SM) is stable. Everywhere else the ground state is a superconductor (SC) which has a fully gapped Bogolioubov de Gennes spectrum and which preserves time reversal symmetry. There is a single superconducting phase, with an order parameter that is given by (\ref{eq: OPallmu}).\label{fig: cleanphasediag}}
\end{figure}

\section{Mean field superconductivity of disordered Dirac fermions}
In this section we discuss the superconductivity of Dirac fermions in the presence of disorder. The section is structured as follows: first, we discuss disorder in the strength of the attraction $g$. We discuss the Harris criterion for determining the relevance of disorder, and show that the disorder is Harris relevant at the critical point $\mu=0$ and $g=g_c$. Thus, any disorder in the attraction strength must change the universality class of the critical point. Next, we consider disorder in the chemical potential $\mu$. We show that disorder in the chemical potential is relevant for $g<g_c$ on the $\mu=0$ line, and that it destroys the semimetal phase (Fig.\ref{fig: cleanphasediag}) by introducing a non-zero density of states. The non-vanishing density of states in turn introduces a weak coupling instability to superconductivity. In the presence of chemical potential disorder, the zero temperature phase diagram in the entire $\mu-g$ plane thus contains a single phase - a superconductor. We are thus driven to the remarkable conclusion that at $\mu=0$ and $g<g_c$, the presence of disorder actually {\it enhances} superconductivity. By introducing a non-vanishing density of states, disorder allows superconductivity to develop in what would have been a semimetallic phase had disorder not intervened. 

The rest of this section is focused on understanding the behavior along the $\mu=0$ line, in the weak coupling regime, in the presence of disorder. We show that for smooth chemical potential disorder, where the disorder is correlated over lengthscales large compared to the superconducting coherence length, the superconducting critical temperature $T_c$ may be extracted from the clean system results by treating the system as being `locally doped.' Meanwhile, for short range correlated chemical potential disorder (which we model using `delta function impurities'), we derive an estimate for $T_c$ based on a `uniform mean field' calculation similar in spirit to the standard Abrikosov-Gorkov theory\cite{Abrikosov-Gorkov} for disordered superconductors. This estimate for $T_c$ is doubly exponentially small in the disorder strength. 

\subsection{Disorder in $g$}
First, we consider static disorder in $g$ i.e. we allow $g$ to be spatially non-uniform, fluctuating about some mean value $\langle g\rangle$. The fluctuations in $g$ are assumed to be independent of time. The only critical point in the phase diagram Fig.\ref{fig: cleanphasediag} that is tuned by $g$ is the critical point on the $\mu=0$ line at $g=g_c$, which was argued to display emergent supersymmetry in the clean system.\cite{grover, Sung-Sik} Whether disorder changes the universality class depends on whether the Harris criterion is satisfied.\cite{Harris} The Harris criterion states that disorder changes the universality class if 
\begin{equation}
\nu d < 2, \label{eq: Harris}
\end{equation}
where $d$ is the spatial dimension and $\nu$ is the critical exponent for the correlation length $\xi$, which diverges near the critical point as $\xi \sim (g-g_c)^{-\nu}$. Intuitively, if the Harris criterion is satisfied, then the typical fluctuation in $g$, averaged over a box of size $\xi^d$, is greater than the remaining distance to the critical point.

For the particular critical point under consideration here, $\nu \cong 3/4$ (to leading order in an $\epsilon$-expansion) and $d=2$.\cite{Sung-Sik2} The Harris criterion is therefore satisfied, and disorder in the interaction strength is a relevant perturbation. It thus follows that disorder in $g$ should change the universality class of the interaction. A determination of the `true' critical point is beyond the scope of this paper. 

\subsection{Disorder in $\mu$}
Static disorder in $\mu$ has an even more dramatic effect: it destroys the semimetal phase by producing a non-vanishing density of states. The Harris criterion is the wrong framework for analyzing the effect of chemical potential disorder. The Harris criterion applies at a critical point which separates regions that are in distinct phases. Meanwhile, the semimetal phase along the $\mu=0$ line separates two regions which are in the {\it same} phase. As a result, all fluctuations about the semimetal phase place us in the same superconducting phase. The relevance of chemical potential disorder was also established using renormalization group arguments in Ref.~\onlinecite{Ludwig}. 
In this paper we will consider two distinct models of disorder: smooth disorder and delta function disorder. We discuss each in turn.

\subsubsection{Smooth disorder} 
The first model we will consider is of smooth disorder. Smooth disorder may be modeled by taking the Hamiltonian (\ref{eq: HBCS}) at $\mu=0$ and adding a term $H_{d}$, where  
\begin{eqnarray}
H_d &=&\int d^2 r \mu_{\vec{r}} \psi^{\dag}_{\vec{r}} \sigma_0 \psi_{\vec{r}}, \\
 \label{eq: smooth disorder}
\langle \mu_{\vec{r}} \rangle &=& 0, \nonumber\\
\langle \mu_{\vec{r}} \mu_{\vec{r'}}\rangle &=& \mu_0^2 \exp\bigg(-\frac{|\vec{r}-\vec{r'}|^2}{2 R^2 }\bigg).
\end{eqnarray}
The correlation length for the disorder $R$ is assumed to be much longer than the superconducting coherence length specified in (\ref{eq: coherence length smooth disorder}). In this limit, the system can be thought of as `locally doped,' and consisting of a network of large electron and hole puddles. 
In each puddle, superconductivity develops as if the system were doped with chemical potential $\mu_0$. In the weak coupling limit, the local order parameter at zero temperature is 
\begin{equation}
\Delta_0 \sim \omega_D \exp(-\pi v^2/ g\mu_0), \label{eq: tcsmoothdisorder}
\end{equation}
which sets a coherence length 
\begin{equation}
\xi = \frac{v}{\Delta} = \frac{v}{\omega_D} \exp(\pi v^2/g\mu_0). \label{eq: coherence length smooth disorder}
\end{equation}
This coherence length must be much smaller than the puddle size $R$ to be in the smooth disorder regime. 

Thus, smooth potential disorder destroys the semimetal phase, replacing it by a network of large electron and hole doped puddles, each of which individually becomes superconducting, with a critical temperature of order (\ref{eq: tcsmoothdisorder}). However, for the sample to be globally superconducting, it is essential that the various electron and hole doped puddles be phase coherent. 

\subsubsection{Phase coherence of locally superconducting puddles}
To estimate the critical temperature for phase coherence of locally superconducting puddles, we consider a specific model for disorder, which takes the form
\begin{equation}
\mu_{\vec{r}} = \mu_0 \sin(x/R) \sin(y/R), \label{eq: specificsmooth}
\end{equation}
where $R$ is much greater than the coherence length (\ref{eq: coherence length smooth disorder}). This has electron and hole doped regions, as well as intermediate regions which are close to undoped. The local coherence length scales as 
\[
\xi_{\vec{r}} = \xi_0^{\mu_{\vec{0}}/\mu_{\vec{r}}} (\omega_D/v)^{-1+\mu_{\vec{0}}/\mu_{\vec{r}}},
\]
where $\xi_0$ is defined by (\ref{eq: coherence length smooth disorder}). Now the region near $x=0$ with $x<\xi(x)$ can be thought of as an undoped `barrier' region separating electron and hole doped islands. From this we conclude that the undoped barrier regions have width $R/\ln(R/\xi_0)$.

Crucially, electron and hole doped puddles enter the same superconducting phase (Fig.\ref{fig: cleanphasediag}), which is mostly spin singlet $\Delta \sim \langle \psi i \sigma_2 \psi \rangle$ for weak disorder $\mu_0 \ll \omega_D$. The Josephson coupling between regions of size $R \times R$ and local order parameter $\Delta_0$, separated by a barrier region of width $W$ is\cite{gonzalez}
\begin{equation}
J = \Delta_0 \frac{\xi_0 R^2}{ W^3}. \label{eq: Josephson1}
\end{equation}
Global phase coherence survives up to temperatures of order $T\approx J$. Substituting $W \approx R/\ln(R/\xi_0)$ into the above equation, we obtain an estimate for the critical temperature for global phase coherence, 
\begin{equation}
T_c \approx \Delta_0 \frac{\xi_0 \ln^3 (R/\xi_0)}{R},\label{eq: Josephson1}
\end{equation}
where $\Delta_0$ is defined by (\ref{eq: tcsmoothdisorder}). Thus, global superconductivity is weaker than local superconductivity by the small parameter $\xi_0/R \ll 1$. 

Note that the global $T_c$ increases as $R$ is made smaller, and appears to diverge as $R\rightarrow 0$. However, the analysis is only valid in the smooth disorder regime $\xi_0/R\ll1$ (and also in the weak disorder regime $\mu_0 \ll \omega_D$). Thus, the global $T_c$ is always smaller than the local puddle $T_c$ by the small parameter $\xi_0/R \ll 1$. 

\subsubsection{Delta-function disorder}
In the strict weak coupling limit, $g\rightarrow 0$, the coherence length (\ref{eq: coherence length smooth disorder}) diverges, and disorder cannot be modeled as being smooth. There is therefore a need for a theory of superconductivity in the presence of short range disorder. The simplest possibility is to consider the limit $R\rightarrow 0$ in (\ref{eq: smooth disorder}). This may be modeled by adding to the Hamiltonian (\ref{eq: HBCS}) $N$ randomly placed positive delta function impurities with impurity strength $V$, and an equal number of randomly placed negative delta function impurities. 
\begin{equation}
H = H_\textrm{BCS} + V \sum_i \eta_i \delta(\vec{x}_i), \label{eq: Hdeltafunctions}
\end{equation}
where $\eta_i=\pm1$, and $V$ is the strength of the disorder. 

We assume that the `impurities' are placed at random, so that the disorder concentration in a box of size $L$ is Poisson distributed, 
\begin{equation}
P_L(n_{\pm}) = \frac{1}{\sqrt{2\pi n_0/L^2}}\exp\bigg(-L^2 \frac{(n_{\pm}-n_0/2)^2}{n_0} \bigg). \label{eq: Poisson}
\end{equation}
Here $P_L(n_{\pm})$ is the probability that an $L\times L$ square box contains $\pm$ impurities with a concentration $n_{\pm}$, $n_0$ is the mean total density of impurities. $L$ must be much bigger than the typical distance between impurities, $l_0=1/\sqrt{n_0}$. 

\subsection{Density of states from short range disorder}
In this section we discuss the density of states arising from short range disorder, in the absence of any interaction ($g=0$). The density of states depends strongly on energy. For the clean system, the density of states scales as 
\begin{equation}
\nu_{\text{clean}} = \frac{\varepsilon}{2\pi v^2}. \label{eq: nuclean}
\end{equation}
Meanwhile, an energy scale $\varepsilon$ also sets a wavelength $ \lambda = v/\varepsilon$. An electron with wavelength $\lambda$ `probes' all impurities within a box of size $\lambda \times \lambda$, and sees a local imbalance $\delta n = n_+ - n_- \neq 0$, which give rise to a local chemical potential 
$\mu = V \delta n$ 
%
This in turn enhances the density of states according to 
%
%
\[
\delta \nu =  \frac{V |\delta n| }{2\pi (\hbar v_F)^2}.
\]
%
Now the probability distribution for $\delta n$ is
\begin{eqnarray}
P(\delta n) &=& \int dn P(n_+ = n+\delta n/2) P(n_- = n - \delta n/2)\nonumber\\
&=&\frac{1}{\sqrt{4 \pi n_0/\lambda}} \exp\bigg(-\lambda^2 \frac{\delta n^2}{2 n_0} \bigg),\nonumber
\end{eqnarray}
which can be turned into a probability distribution for the correction to the density of states coming from a local imbalance
\begin{equation}
P( \delta \nu_{\lambda}) = \frac{\lambda 4 \pi }{V \sqrt{2\pi n_0} } \exp\bigg(- \lambda^2 \frac{(2\pi v^2)^2  \delta \nu_{\lambda}^2}{ V^2 2 n_0}\bigg). \label{eq: dosimbalance}
\end{equation}
The root mean square density of states from local imbalance scales as 
\[
\delta \nu_\textrm{rms}^\textrm{imbalance} =  \frac{V\sqrt{n_0}}{\lambda 2\pi v^2 }.
\]
Thus, at the energy scale $\varepsilon = v/\lambda$, the correction to the density of states from local shifts of the chemical potential is 
\[
\delta \nu_\textrm{rms}^\textrm{imbalance}(\varepsilon) =  \frac{V \sqrt{n_0}}{2\pi v^3}  \varepsilon.
\]
This scales with energy in the same way as (\ref{eq: nuclean}). Thus, the `local chemical potential' merely changes the co-efficient in (\ref{eq: nuclean}) and does not qualitatively alter the energy denendence of the density of states. However, there is a second contribution to the density of states, which comes about due to scattering of the electrons. 
%
%

\subsubsection{Density of states from scattering}
The density of states can be extracted from the electron Green function according to 
\[
\nu(\omega) = -\frac{1}{\pi} \Tr \Im G (\omega), 
\]
where $G$ is the retarded Green function. 
Now, the Green function for a Dirac fermion at $\mu=0$, after ensemble averaging over disorder takes the form
\[
\langle \vec{k} | G(\omega)|\vec{k}\rangle = \frac{1}{\omega - v_F k + i/\tau},
\]
where the scattering time $\tau$ may be estimated using the self consistent Born approximation (SCBA) as in Ref.~\onlinecite{Ando}. This gives rise to a density of states equal to 
\begin{equation}
\nu_\textrm{SCBA}(\omega \rightarrow 0) = \frac{\Lambda}{ n_0 V^2}\exp\bigg(-\frac{v^2}{n_0 V^2}\bigg), \label{eq: dosscba}
\end{equation}
where $\Lambda$ is an ultraviolet cutoff of order the electronic bandwidth and $n_0$ is the mean density of impurities. Note that there is a non-vanishing density of states even at zero energy. Thus, scattering on disorder destroys the semimetal phase, in agreement with Ref.~\onlinecite{Ludwig}. The SCBA applies in the weak disorder limit $n_0 V^2/v^2 \ll 1$. In the strong disorder limit the zero energy density of states scales linearly with impurity concentration, and can be thought of as coming from low energy bound states. In this paper we will focus on the weak disorder limit.  

\subsection{Superconductivity from short range disorder}
We observed that in the presence of short range disorder, the system develops a non-vanishing density of states. In the presence of a non-vanishing density of states, there is a weak coupling instability to superconductivity. In this section, we search for a spatially uniform superconducting phase in the weakly disordered Dirac fermion system. 

We have a disordered system of Dirac fermions. The exact single particle eigenstates of the disordered system are $|\psi_{\alpha}\rangle$. The disorder is time reversal preserving. Thus, Kramer's theorem holds and all states come in Kramer's doublets $|\psi_{\alpha,\sigma}\rangle$, where $\sigma$ is a Kramer's index. Let $\vec{\tau}$ be Pauli matrices acting in the space of the Kramer's doublet. 
The BCS Hamiltonian can then be written as 
\[
H = \sum_{\alpha} \varepsilon_{\alpha} \psi^{\dag}_\alpha \tau_0 \psi_{\alpha} - g_{\alpha \beta} (\psi^{\dag}_{\alpha} i \tau_2 \psi_{\alpha}^*)(\psi_{\beta}^T i \tau_2 \psi_{\beta}),
\]
where repeated indices are summed over. As usual, we have projected the interaction onto the BCS channel. We introduce a pairing field
\[
\Delta_{\beta} =  \sum_{\alpha} g_{\beta \alpha} \langle \psi_{\alpha}^T i \sigma_2 \psi_{\alpha} \rangle,
\]
by going to the path integral and using a Hubbard Stratanovich transformation. After decoupling the four fermion interaction, we obtain a Lagrangian of the form
\[
L = \psi^{\dag}_{\alpha} (i \omega_n - \varepsilon_\alpha ) \sigma_0 \psi_{\alpha} + \Delta^*_{\alpha} \psi_{\alpha} i \sigma_2 \psi_{\alpha} +\textrm{c.c.}  +\sum_{\alpha \beta}  \frac{\Delta_{\alpha}^* g^{-1}_{\alpha \beta} \Delta_{\beta}}{2},
\]
where $\omega_n$ is a fermionic Matsubara frequency. Upon going to the Nambu spinor basis and integrating out the fermions exactly, we obtain an action purely for the order parameter fields, which takes the form
\[
L = - \Tr \ln \big[\omega_n^2 + \varepsilon_{\alpha}^2 + |\Delta_{\alpha}|^2\big] +\sum_{\alpha \beta}  \frac{\Delta_{\alpha}^* g^{-1}_{\alpha \beta} \Delta_{\beta}}{2}.
\]
From this we obtain the gap equation, which after integration over Matsubara frequencies (at zero temperature) takes the form
\begin{equation}
\sum_{\alpha} \frac{ \Delta_{\alpha}}{\sqrt{\varepsilon_{\alpha}^2+\Delta_{\alpha}^2}} = \sum_{\beta} g^{-1}_{\alpha \beta} \Delta_{\beta}. \label{eq: exact}
\end{equation}
Note that we have not made any approximations in deriving (\ref{eq: exact}) (except for projecting the interaction on the BCS channel). So far, everything is exact, for a given realization of disorder. 

Now we ensemble average over disorder. After ensemble averaging over disorder, translation invariance is restored, and the eigenstates are indexed by momentum. The interaction is a constant in momentum space, so that the gap equation takes the form 
\begin{equation}
\int \frac{ \nu(\varepsilon) d\varepsilon }{\sqrt{\varepsilon^2+\Delta_{\alpha}^2}} = \frac{1}{\tilde g}, \label{eq: AbrikosovGorkov}
\end{equation}
where $\nu$ is the disorder averaged density of states and $\tilde g$ is the disorder averaged interaction. This `disorder averaged gap equation' is the naive Dirac fermion analog of the `Abrikosov-Gorkov' theory for superconductivity in disordered metals.\cite{Abrikosov-Gorkov} Now, the vertex correction to $g$ arising from disorder at the one loop level is convergent, so that disorder does not produce a singular renormalization of $g$. We have $\tilde g = A g$, where $A$ is some $O(1)$ prefactor. We drop this prefactor for compactness, and use $\tilde g = g$. Substituting (\ref{eq: dosscba}) into (\ref{eq: AbrikosovGorkov}) and solving, we obtain 
\begin{align}
T_c \approx \Delta &\sim \omega_D \exp \left(-\frac{\tau}{g} \right)\nonumber\\
& \sim \omega_D \exp\left[ - \frac{n_0 V^2}{g v} \exp\left( \frac{v^2}{n_0 V^2}\right) \right], \label{eq: TcAG}
\end{align}
where the scattering time $\tau$ is exponentially sensitive to disorder strength. Note that there is a non-zero $T_c$ for any value of $g$, however small. However, the critical temperature is {\it doubly} exponentially small in the disorder strength. As we shall see later in the paper, (\ref{eq: TcAG}) is a gross underestimate of the strength of the superconducting instability in the disordered system. The true $T_c$ is actually only exponentially small in weak disorder, not doubly exponentially small. 
However, we stress that $T_c$ is non-zero even in (\ref{eq: TcAG}) i.e. even though the clean system does not superconduct, the introduction of weak short range disorder introduces a weak coupling instability to superconductivity. This, disorder has the surprising effect of enabling superconductivity, by destroying the semimetal phase. 

\section{Superconductivity from rare puddles}
In this section, we investigate the possibility that superconductivity may actually develop at a temperature much higher than (\ref{eq: TcAG}), because of mesoscopic fluctuation effects that are ignored in the uniform mean field analysis. The analysis in this section is inspired by the work\cite{Oreta, Spivak, Lamacraft} on superconductor to metal transitions in the presence of disorder, but with important differences arising from the different nature of the order parameter, and the fact that the `critical point' $\mu=0$ now separates two regions in the same phase rather than two regions in different phases. 

The specific possibility that we investigate is the following: in a sample where disorder is weak and the density of states is small, there may nonetheless be regions where disorder is stronger, and the local density of states is larger. These regions will have strongly enhanced local superconductivity. Even small fluctuations in the disorder concentration will have large effects on the local $T_c$, because of the double exponential sensitivity of $T_c$ to disorder concentration (\ref{eq: TcAG}). Thus, one expects that in any disordered sample, there will be regions where the local $T_c$ is much higher than (\ref{eq: TcAG}). If the Josephson coupling between these locally superconducting regions is strong enough to establish global phase coherence, then the sample will superconduct. The critical temperature for this `puddle based superconductivity' will be the temperature at which the locally superconducting regions lose phase coherence. As we will show, this temperature is only exponentially sensitive to disorder concentration, unlike the double exponential dependence in (\ref{eq: TcAG}). Thus, it represents a strong enhancement over (\ref{eq: TcAG}). 

This section is structured as follows: first we discuss the probability that a region of size $L$ is locally superconducting, and find the most likely size of superconducting puddle $L_*$. Then we discuss the Josephson coupling between superconducting regions, and the temperature at which global phase coherence is established. The discussion assumes that disorder is weak and dilute i.e. the sample is `almost clean.'   We conclude by making a few comments on the `gauge glass' behavior that arises when external magnetic field is applied. 

\begin{figure}
\includegraphics[width = \columnwidth]{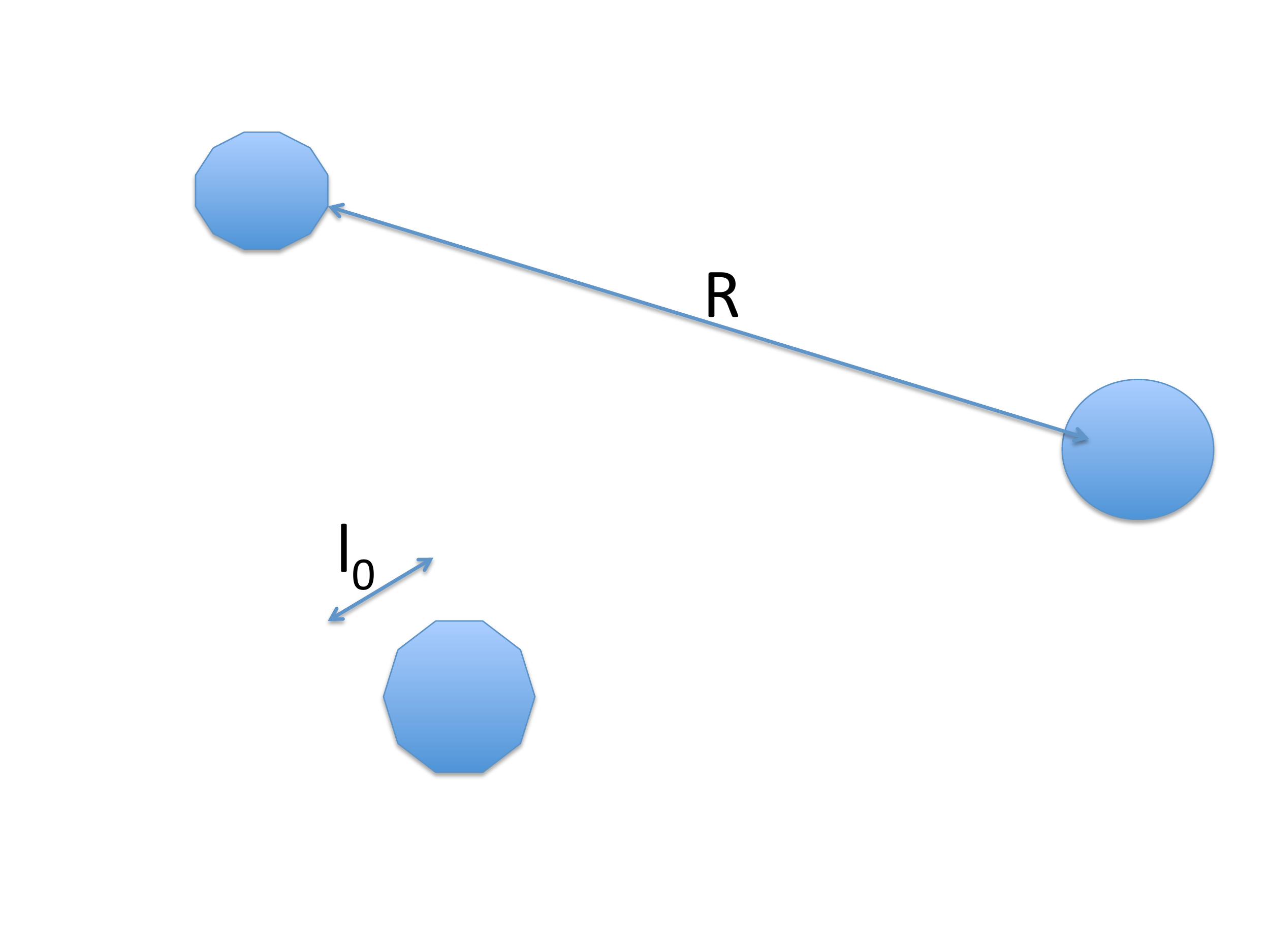}
\caption{Schematic illustration of puddle based superconductivity. There are rare puddles which develop superconductivity, and Josephson coupling then establishes phase coherence between these puddles \label{fig: puddles}}. 
\end{figure}

\subsection{Optimum size of superconducting puddles}
What is the probability that a puddle of a given size $L$ displays local superconductivity at a given temperature $T$? Maximising this probability by varying $L$ will tell us the optimal size of the superconducting puddle. 

We assume that the following equation is true
\[
T_c(\vec{r}) \approx \omega_D \exp\big(-1/g\nu(\vec{r})\big),
\]
where $T_c(\vec{r})$ is the local $T_c$ in a region of size of order the coherence length, and $\nu(\vec{r})$ is the local density of states in this region. A puddle of size $L$ will be locally superconducting if and only if the following two conditions are satisfied: (i) The local $T_c \ge T$ and (ii) the puddle is bigger than the local coherence length, $L \ge (\hbar v_F/\omega_D) \exp(1/g\nu)$. 

We can reformulate this as a condition on the density of states of the puddle. A puddle of size $L$ will be locally superconducting IFF the density of states per unit area on the puddle 
\begin{equation}
\nu \ge \nu_\textrm{min} = \max\left(\frac{1}{g \ln \omega_D/T}, \frac{1}{g \ln (\omega_D L/\hbar v_F}\right). \label{eq: numin}
\end{equation} 
The probability that the puddle is superconducting is given by $\int_{\nu_\textrm{min}}^{\infty} d\nu P(\nu)$. 

It is useful to define the thermal length 
\begin{equation}
L_T = v/T. \label{eq: thermal length}
\end{equation}
For puddles bigger than the thermal length, the temperature is the key cutoff,
\begin{equation}
P_\textrm{SC}(L>L_T) = \int_{1/(g \ln L_T/a)}^{\infty} d\nu\,P(\nu). \label{eq: lt}
\end{equation}
Meanwhile, for puddles smaller than the thermal length, the puddle size is the key cutoff,
\begin{equation}
P_\textrm{SC}(L<L_T) = \int_{1/(g \ln L/ a)}^{\infty} d\nu\,P(\nu). \label{eq: l}
\end{equation}
%
Now, the density of states arises from disorder, and we can re-express
\begin{equation}
P_\textrm{SC}(L) = \int_{\nu_{min}}^{\infty} d\nu\,P(\nu) = \int_{n_\textrm{min}}^{\infty} dn\,P(n), \label{eq: intovern}
\end{equation}
where $\nu_\textrm{min} = \frac{1}{g}\max(\frac{1}{\ln L_T/a}, \frac{1}{\ln L/a})$, and $\nu$ is related to the local disorder concentration $n$ by Eq.~(\ref{eq: dosscba}). In the weak disorder regime, Eq.~(\ref{eq: dosscba}) tells us that $\nu_\textrm{min} $ sets 
\begin{equation}
n_\textrm{min} = \frac{v^2}{V^2\ln(\Lambda/n_0V^2 \nu_\textrm{min})}. \label{eq: nmin}
\end{equation}
Meanwhile, $P(n)$ is given by 
\begin{equation}
P_L(n) = \frac{1}{\sqrt{2\pi n_0/L^2}}\exp\bigg(-L^2 \frac{(n-n_0)^2}{n_0} \bigg). \label{eq: Poisson2}
\end{equation}
It is easier to integrate over $P(n)$ since this distribution is just Gaussian. 
Upon substitution into (\ref{eq: l}) we find that the probability of superconductivity is 
\[
P_\textrm{SC}(L) = \frac{1}{2}\textrm{erfc}\bigg(\frac{L(n_\textrm{min}-n_0)}{\sqrt{n_0}} \bigg),
\]
where we assume $n_\textrm{min} \gg n_0$ otherwise we are just dealing with uniform mean field superconductivity. Now $n_{\min}$ is given by (\ref{eq: nmin}) and $\nu_{\min}$ is given by (\ref{eq: numin}). Thus, $n_{\min}$  has a log(log) dependence on $L$. Substituting into the above equation and plotting, we find that $P_L(SC)$ is a monotonically decreasing function of puddle size over the entire range of sizes satisfying $n_0 L^2 \gg 1$, with smaller puddles being exponentially more likely to be superconducting.

Thus, smaller sized puddles are much more likely to superconduct. The smallest sized puddle that it is meaningful to talk about is a puddle with size of order $l_0\approx1/\sqrt{n_0}$, where $l_0$ is the typical spacing between impurities (which is assumed to be large in the weak disorder regime). Thus, the most probable superconducting puddle has a size of order $l_0$. The local $T_c$ for this puddle may be determined by remembering that this puddle has a local coherence length (at zero temperature) of order $l_0$. Thus $v/\Delta \approx v/T_c \approx l_0$. This gives rise to a local critical temperature
\begin{equation}
T_c^\textrm{loc} \sim v \sqrt{n_0} .\label{eq: Tcloc}
\end{equation}
In the weak disorder regime $n_0 \rightarrow 0$ this is small, but it is only linearly small in weak disorder, not doubly exponentially small. Thus, the local $T_c$ for the superconducting puddles is enormously enhanced over the uniform mean field $T_c$. 

The probability that a given region of size $l_0$ superconducts is 
\begin{eqnarray}
P_\textrm{SC}(l_0) &\approx& \text{erfc}\bigg(\frac{l_0(n_{min}-n_0)}{\sqrt{n_0}} \bigg)\nonumber\\
&=&  \text{erfc}\bigg(\frac{v^2}{n_0 V^2 \ln \big(\frac{g v \ln(\omega_D/v \sqrt{n_0})}{n_0V^2} \big)} -1\bigg).\nonumber
\end{eqnarray}
Now $v^2/n_0V^2 \gg 1$ to be in the strong disorder regime, so the above probability is much less than one. We can approximate it as 
\[
P_\textrm{SC}(l_0) \sim \exp\bigg(-\frac{v^4}{n_0^2 V^2 \ln^2 \big(\frac{g v}{n_0V^2} \big)}\bigg),
\]
where we have neglected double log terms. The typical separation between superconducting puddles is then $R$, where 
\begin{equation}
R \sim l_0 \exp\bigg(\frac12 \frac{v^4}{n_0^2 V^2 \ln^2 \big(\frac{g v}{n_0V^2} \big)}\bigg) \gg l_0. \label{eq: interpuddle distance}
\end{equation}
In order for the sample to be globally superconducting, the Josephson coupling between distant puddles must be strong enough to establish phase coherence. 
\subsection{Josephson coupling between distant puddles}
In order for the puddles to be phase coherent, the thermal length must be larger than the typical inter-puddle spacing, otherwise thermal decoherence will destroy the Josephson coupling. This sets a bound on the temperature for global phase coherence. 
\begin{eqnarray}
T_\textrm{KT} &\le& v\sqrt{n_0} \exp\bigg(-\frac12 \frac{v^4}{n_0^2 V^2 \ln^2 \big(\frac{g v}{n_0V^2} \big)}\bigg) \nonumber\\ 
&=& T_c^\textrm{loc} \exp\bigg(-\frac12 \frac{v^4}{n_0^2 V^2 \ln^2 \big(\frac{g v}{n_0V^2} \big)}\bigg). \label{eq: thermal bound}
\end{eqnarray}
This is exponentially smaller than the local $T_c$ (\ref{eq: Tcloc}), but note that it is still only exponentially small in weak disorder, not doubly exponentially small. 

As long as the bound (\ref{eq: thermal bound}) is satisfied, we can model the Josephson coupling between puddles using the zero temperature results obtained by Ref.~\onlinecite{gonzalez}. In Ref.~\onlinecite{gonzalez}, it was determined that the Josephson coupling between distant puddles at zero temperature scales as 
\[
J = \frac{v W^2}{R^3},
\]
where $W\approx l_0$ is the size of the puddle and $R$ is the typical inter-puddle separation (\ref{eq: interpuddle distance}), and the intervening region is modelled as being at the Dirac point. Note that the Josephson coupling is power law with distance, and hence long ranged. Taking into account the finite density of states in the intervening region will alter the power, making the Josephson coupling decay more slowly with distance, but we use the above result to be conservative. Taking into account the finite density of states in the intervening region will just add an $O(1)$ prefactor to the exponent in (\ref{eq: TKT}) (and will enhance the critical temperature). 

Note that since the superconductivity is mostly $s$-wave close to the Dirac point, the Josephson coupling is {\it unfrustrated}. This is a major difference to Ref.~\onlinecite{Oreta, Spivak, Lamacraft}, where frustration arising from the d-wave nature of the order parameter dramatically impacted the physics. 

The unfrustrated Josephson coupling will be strong enough to establish phase coherence upto a temperature $T_\textrm{KT} \sim J$. This tells us that the maximum temperature up to which global phase coherence can be expected to occur is 
\begin{eqnarray}
T_\textrm{KT} &=& v\sqrt{n_0} \exp\bigg(-\frac32 \frac{v^4}{n_0^2 V^2 \ln^2 \big(\frac{g v}{n_0V^2} \big)}\bigg) \nonumber\\ 
&=& T_c^\textrm{loc} \exp\bigg(-\frac32 \frac{v^4}{n_0^2 V^2 \ln^2 \big(\frac{g v}{n_0V^2} \big)}\bigg). \label{eq: TKT}
\end{eqnarray}
This is a smaller temperature than the thermal bound (\ref{eq: thermal bound}), so we conclude that this is the true critical temperature at which global phase coherence is lost. Note that this temperature is only exponentially small in weak disorder, not doubly exponentially small as in (\ref{eq: TcAG}). Thus, the critical temperature arising from rare superconducting puddles with phase coherence is enormously higher than the critical temperature for uniform superconductivity. 
\subsection{Magnetic field and gauge glass behavior}
In the above discussion, the Josephson coupling between superconducting puddles was unfrustrated, because each puddle is mostly `$s$-wave.' The application of a transverse magnetic field frustrates the Josephson coupling, introducing a random phase difference $\int \vec{A} \cdot \vec{dr}$ to each Josephson link, where $\vec{A}$ is the magnetic vector potential and the integral goes along the line connecting two puddles. At zero temperature, this turns the globally phase coherence superconductor into a `gauge glass'.\cite{Fiegelman, MPAF} At finite temperature, vortex creep (i.e., phase slips across the Josephson junctions) will introduce a non-zero resistance. Increasing temperature at non-zero magnetic field then drives a smooth crossover to the high temperature semimetallic phase. 

\subsection{Fate of the triplet pairing mode}
In Section II, we showed that in the clean system away from charge neutrality $\mu \neq 0$, s-wave pairing induces spin triplet `p-wave' pairing, with the p-wave amplitude being small when $\mu \ll \omega_D$. Insofar as a particular puddle has $\mu \neq 0$, s-wave pairing will also induce some local p-wave component of the order parameter. However, this p-wave component will be small, and moreover the Josephson couplings between p-wave pieces will be strongly frustrated, and so we do not expect any long range ordered p-wave component to the order parameter.

\section{Replica Renormalization group}

In this section, we examine the interplay of disorder and superconductivity using a perturbative renormalization group (RG) treatment. This technique is appropriate for studying what happens close to the supersymmetric critical point $\mu=0$, $g=g_c$. The clean critical point is perturbatively accessible in the $\epsilon$-expansion\cite{ScottThomas,Sung-Sik2} close to four spacetime dimensions. To take the disorder into account we use the replica trick.\cite{sachdev} We first show in Sec.~\ref{sec:mudisRG} using a one-loop RG analysis that chemical potential disorder is perturbatively irrelevant at the supersymmetric critical point. However, this result is deceptive because chemical potential disorder generates disorder in the BCS coupling at the two-loop level, and this type of disorder turns out be relevant at the critical point as shown in Sec.~\ref{sec:BCSdisRG}.

\subsection{Chemical potential disorder}
\label{sec:mudisRG}

\begin{figure}[t]
\begin{center}
\includegraphics[width=\columnwidth]{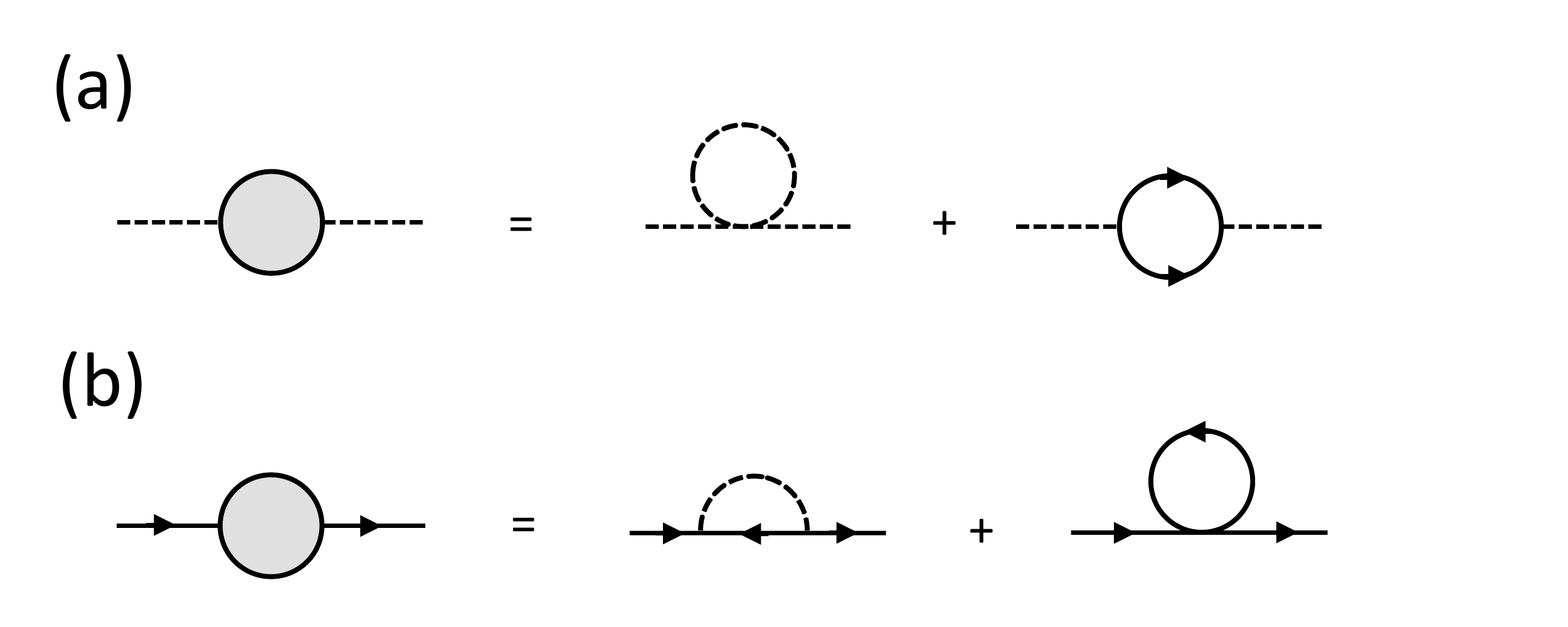}
\end{center}
\caption{One-loop diagrams in the theory with chemical potential disorder for (a) boson mass, field and velocity renormalization and (b) fermion field and velocity renormalization. The dashed lines represent boson propagators and the solid lines, fermion propagators. The $\phi^4$ coupling $\lambda$ is represented by four dashed lines meeting at a point, and the disorder-induced coupling $\Delta_\mu$ is represented by four solid lines meeting at a point. A dashed line ending on a solid line corresponds to the boson-fermion coupling $h$.}
\label{fig:FieldRenFer}
\end{figure}
\begin{figure}[t]
\begin{center}
\includegraphics[width=\columnwidth]{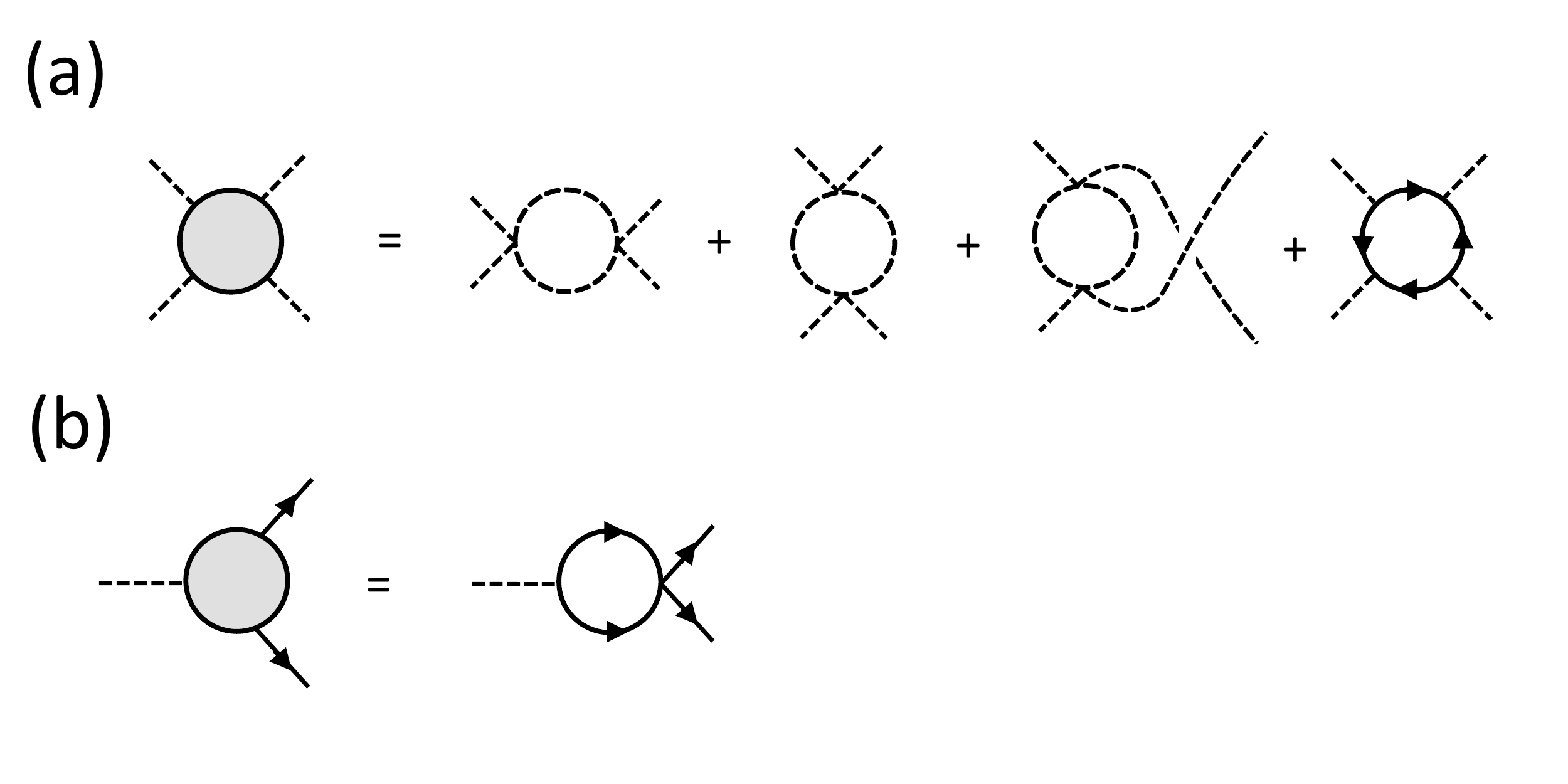}
\end{center}
\caption{One-loop diagrams in the theory with chemical potential disorder for renormalization of (a) the 4-boson coupling $\lambda$ and (b) the boson-fermion coupling $h$.}
\label{fig:VertexRenFer}
\end{figure}
\begin{figure}[t]
\begin{center}
\includegraphics[width=\columnwidth]{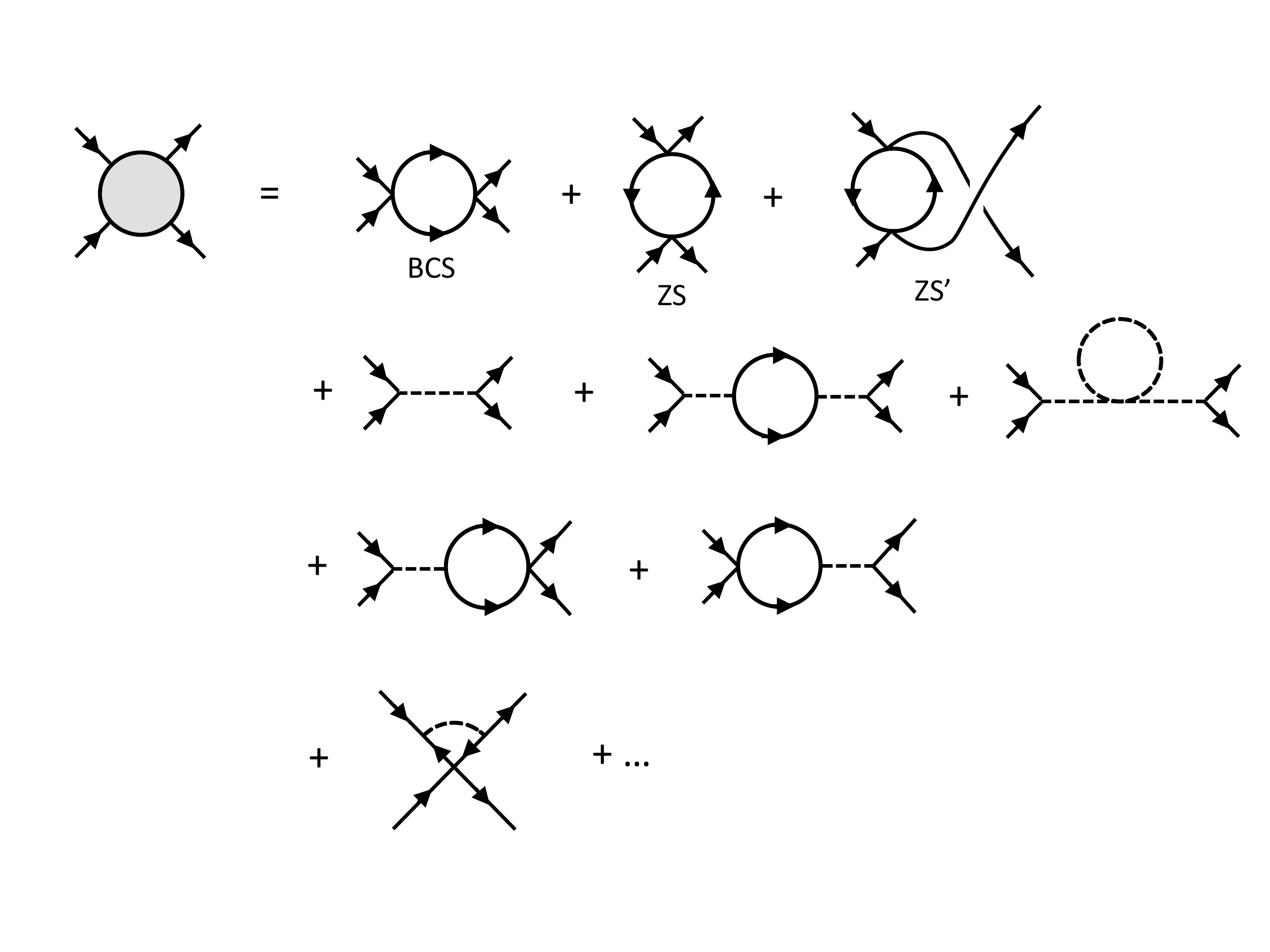}
\end{center}
\caption{One-loop diagrams in the theory with chemical potential disorder for renormalization of the disorder strength $\Delta_\mu$.}
\label{fig:DeltaRenFer}
\end{figure}
We consider the problem of Dirac fermions $\psi$ coupled to the bosonic $s$-wave superconducting order parameter $\phi\equiv\Delta_s$. The Euclidean Lagrangian is\cite{grover,Sung-Sik}
\begin{align}
\mathcal{L}&=i\overline{\psi}(\gamma_0\partial_\tau+c_f\gamma_i\partial_i)\psi
+\frac{1}{2}\left(|\partial_\tau\phi|^2+c_b^2|\partial_i\phi|^2\right)\nonumber\\
&+\frac{r}{2}|\phi|^2
+\frac{\lambda}{4!}|\phi|^4
+h\left(\phi^*\psi^Ti\gamma_2\psi+\mathrm{h.c.}\right)
+\mu(\mathbf{x})\psi^\dag\psi,\nonumber
\end{align}
where $\psi=\left(\begin{array}{cc}
\psi_{\uparrow} & \psi_{\downarrow}\end{array}\right)^T$ is a two-component spinor, $c_f$ is the fermion velocity, $c_b$ is the boson velocity, $\lambda>0$ is a contact 4-boson interaction, $h$ is a boson-fermion coupling, and $r\sim (g_c-g)$ is a parameter which drives the quantum phase transition between the Dirac semimetal ($r>0$) and the superconductor ($r<0$). The Dirac conjugate is $\overline{\psi}=-i\psi^\dag\gamma_0$ where we choose the Dirac matrices to be $\gamma_0=\sigma_3$, $\gamma_1=\sigma_1$, and $\gamma_2=\sigma_2$, where $\sigma_i$, $i=1,2,3$ are the Pauli matrices. Short-range chemical potential disorder is modeled by a random variable $\mu(\mathbf{x})$ with Gaussian distribution centered at zero,
\begin{align}
P[\mu(\mathbf{x})]\propto e^{-\int d^dx\,\mu(\mathbf{x})^2/2\Delta_\mu},\nonumber
\end{align}
where $\Delta_\mu\sim n_0 V^2$ is the disorder strength. The disorder can be integrated out using the replica trick.\cite{sachdev} The replicated action takes the form
\begin{align}\label{Sreplicated}
S=S_f+S_b+S_{bf}+S_\textrm{dis},
\end{align}
with
\begin{align}
S_f&=\sum_{a=1}^n\int d^Dx\,i\overline{\psi}_a(\gamma_0\partial_\tau
+c_f\gamma_i\partial_i)\psi_a,\nonumber\\
S_b&=\sum_{a=1}^n\int d^Dx\left(\frac{1}{2}\left(|\partial_\tau\phi_a|^2
+c_b^2|\partial_i\phi_a|^2\right)\right.\nonumber\\
&\hspace{20mm}\left.+\frac{r}{2}|\phi_a|^2
+\frac{\lambda}{4!}|\phi_a|^4\right),\nonumber\\
S_{bf}&=\sum_{a=1}^n\int d^Dx\,h(\phi_a^*\psi_a^Ti\gamma_2\psi_a+\mathrm{h.c.}),
\nonumber\\
S_\textrm{dis}&=-\frac{\Delta_\mu}{2}\sum_{a,b=1}^n
\int d^dx\int d\tau\int d\tau'\nonumber\\
&\hspace{15mm}\times(\psi_a^\dag\psi_a)(\mathbf{x},\tau)
(\psi_b^\dag\psi_b)(\mathbf{x},\tau'),\label{SdisFer}
\end{align}
where $n$ is the replica index to be set to zero at the end of the calculation. We denote the number of spacetime dimensions by $D=4-\epsilon$ and the number of spatial dimensions by $d=D-1$. The clean supersymmetric critical point is accessible with a one-loop calculation,\cite{ScottThomas,Sung-Sik2} hence we perform a one-loop calculation including disorder. The one-loop diagrams for the boson and fermion two-point functions are given in Fig.~\ref{fig:FieldRenFer}, and those for the four-boson vertex are given in Fig.~\ref{fig:VertexRenFer}(a), the unique one-loop diagram for the boson-fermion vertex is given in Fig.~\ref{fig:VertexRenFer}(b), and the one-loop diagrams for the disorder-induced four-fermion vertex are given in Fig.~\ref{fig:DeltaRenFer}. The lack of Lorentz invariance in the bare theory ($c_f\neq c_b$) leads to anisotropic momentum integrals which are handled using an approach similar to that of Ref.~\onlinecite{vojta2000}. We write the spacetime momentum as $(p_0,\mathbf{p})=p\mathbf{n}$ where $p$ is the magnitude $p=\sqrt{p_0^2+\mathbf{p}^2}$ and $\mathbf{n}$ is a unit vector. The angular integral $\int d\Omega_\mathbf{n}$ only contains information about the anisotropy $c_f\neq c_b$ and does not diverge. We evaluate it in four dimensions, i.e., over the unit three-sphere. The integral over the magnitude $p$ typically diverges in four dimensions and is evaluated in $D=4-\epsilon$ dimensions. In Fig.~\ref{fig:DeltaRenFer}, only the first three diagrams contribute. The remaining diagrams cannot generate an effective interaction which is nonlocal in time, and therefore do not renormalize $\Delta_\mu$. On the critical hypersurface $r=0$, the one-loop RG equations are obtained as follows,
\begin{align}
\frac{dc_f}{d\ell}&=\frac{32h^2(c_b-c_f)}{3c_b(c_b+c_f)^2}-\frac{\Delta_\mu}{c_f},\nn
\frac{dc_b}{d\ell}&=-\frac{2h^2(c_b^2-c_f^2)}{c_bc_f^3},\nn
\frac{d\lambda}{d\ell}&=\left(\epsilon-\frac{8h^2}{c_f^3}\right)\lambda-\frac{5\lambda^2}{3c_b^3}+\frac{192h^4}{c_f^3},\nn
\frac{dh^2}{d\ell}&=\epsilon h^2-\left(\frac{1}{c_f^3}+\frac{8}{c_b(c_b+c_f)^2}\right)4h^4,\nn
\frac{d\Delta_\mu}{d\ell}&=\left(-1+\epsilon-\frac{32h^2}{c_b(c_b+c_f)^2}\right)\Delta_\mu,\nonumber
\end{align}
to $\mathcal{O}(\epsilon)$. We search for the clean supersymmetric critical point and linearize the RG equations around that fixed point. We indeed find a clean fixed point ($\Delta_\mu^*=0$) with emergent Lorentz invariance ($c_f^*=c_b^*=c=1$) and emergent $\mathcal{N}=2$ supersymmetry ($(h^2)^*=\frac{\epsilon}{12}$, $\lambda^*=\epsilon$) corresponding to the Wess-Zumino model with one chiral multiplet.\cite{Sung-Sik} A study of the full RG equations including the flow of $r$ shows that the fixed point is indeed at $r^*=0$. Linearizing the RG equations at the critical point, we find one relevant direction, one marginal direction, and four irrelevant directions. The relevant direction corresponds to the mass parameter $r$ with eigenvalue $y_r=2-\epsilon+\mathcal{O}(\epsilon^2)$, which gives an order parameter exponent $\nu=\frac{1}{2}+\frac{\epsilon}{4}+\mathcal{O}(\epsilon^2)$ in agreement with Ref.~\onlinecite{ScottThomas}. This is the direction which drives the transition. The difference of fermion and boson velocities $c_f-c_b$ as well as the couplings $\lambda$ and $h^2$ have a nonzero projection only along irrelevant directions. (The sum of fermion and boson velocities $c_f+c_b$ has a nonzero projection along the marginal direction.) The only direction along which the disorder strength $\Delta_\mu$ has a nonzero projection is also irrelevant, with eigenvalue
\begin{align}\label{eigDmu}
y_{\Delta_\mu}=-1+\frac{\epsilon}{3}+\mathcal{O}(\epsilon^2),
\end{align}
which is negative and thus irrelevant for small $\epsilon$, and even in the limit $\epsilon\rightarrow 1$ corresponding to the physical case of 2+1 dimensions (although corrections of $\mathcal{O}(\epsilon^2)$ cannot be neglected in this case). Therefore disorder in the chemical potential appears to be an irrelevant perturbation at the supersymmetric critical point.

\subsection{Disorder in the BCS coupling}
\label{sec:BCSdisRG}

The analysis in the previous section could lead us to believe that the supersymmetric critical point is stable against disorder in the chemical potential. However, disorder in the chemical potential will induce randomness in the coefficient of the $|\Delta_s|^2$ term in the Landau-Ginzburg action (\ref{eq: bigLagrangian}) as can be seen from the presence of a $\mu^2|\Delta_s|^2$ term in the clean case. In other words, an interaction of the form (\ref{SdisFer}) but for the bosonic order parameter $\phi$ will be generated at two loops, with a coefficient $\Delta_V\propto h^4\Delta_\mu$. This was missed in our one-loop calculation for chemical potential disorder. However, this interaction is perturbatively relevant at the Gaussian fixed point and should be included in the calculation. More generally, randomness in the BCS coupling $g$ also gives rise to a random coefficient for the $|\Delta_s|^2$ term. As a result, at the critical point, chemical potential disorder is a dangerous irrelevant perturbation, which generates a relevant four-boson term. We therefore repeat the one-loop RG analysis but replace Eq.~(\ref{SdisFer}) by this four-boson term. The Euclidean action again takes the form (\ref{Sreplicated}), but with $S_\textrm{dis}$ given by
\begin{align}
&S_\textrm{dis}=-\frac{\Delta_V}{2}\sum_{a,b=1}^n
\int d^dx\int d\tau\int d\tau'\nonumber\\
&\hspace{15mm}\times(\phi_a^*\phi_a)(\mathbf{x},\tau)
(\phi_b^*\phi_b)(\mathbf{x},\tau'),\nonumber
\end{align}
where $\Delta_V\propto h^4\Delta_\mu\sim n_0V^2$ is the disorder strength.
\begin{figure}[t]
\begin{center}
\includegraphics[width=\columnwidth]{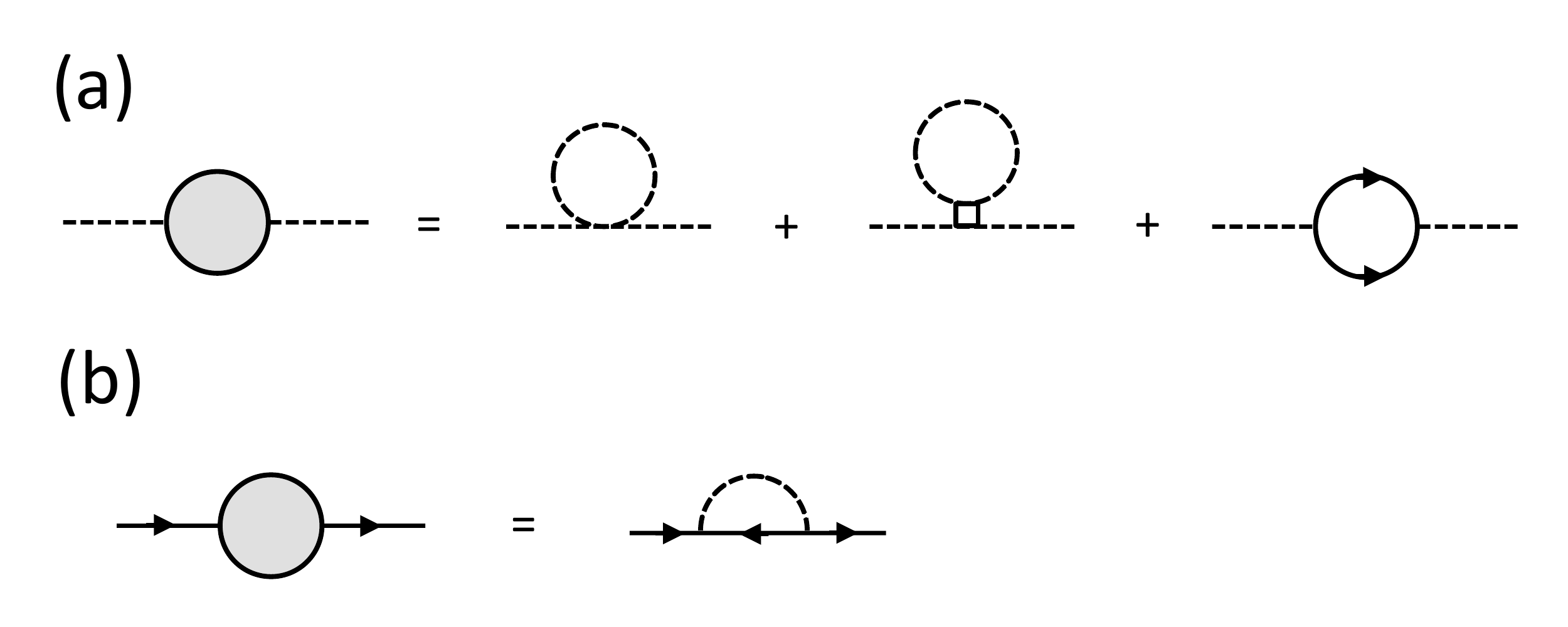}
\end{center}
\caption{One-loop diagrams in the theory with boson mass disorder for (a) boson mass, field and velocity renormalization and (b) fermion field and velocity renormalization. The disorder-induced four-boson coupling $\Delta_V$ is represented by a square box.}
\label{fig:FieldRen2}
\end{figure}
\begin{figure}[t]
\begin{center}
\includegraphics[width=\columnwidth]{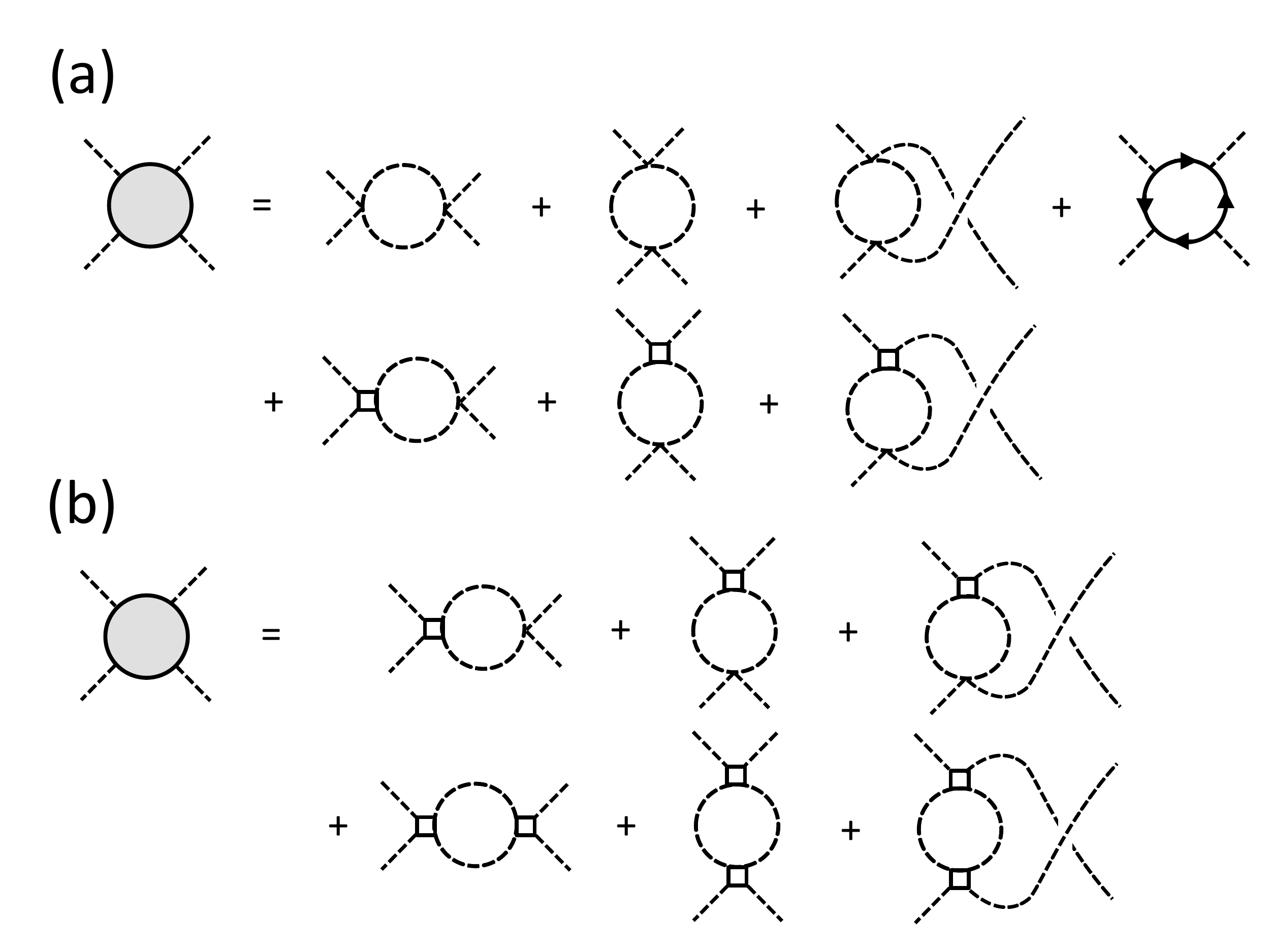}
\end{center}
\caption{One-loop diagrams in the theory with boson mass disorder for renormalization of (a) the 4-boson coupling $\lambda$ and (b) the disorder strength $\Delta_V$. In this case there is no renormalization of the boson-fermion vertex at one loop.}
\label{fig:VertexRen2}
\end{figure}
The one-loop diagrams for the boson and fermion two-point functions are given in Fig.~\ref{fig:FieldRen2}, and those for the boson four-point functions are given in Fig.~\ref{fig:VertexRen2}. Four-fermion interactions will be generated under the RG but are irrelevant for small $\epsilon$. In contrast with the time-reversal symmetry breaking transition for Dirac fermions where the order parameter is in the particle-hole channel,\cite{vojta2000,xu2010} here there is no renormalization of the boson-fermion vertex at one loop. On the critical hypersurface $r=0$, the one-loop RG equations are
\begin{align}
\frac{dc_f}{d\ell}&=\frac{32h^2(c_b-c_f)}{3c_b(c_b+c_f)^2},\nn
\frac{dc_b}{d\ell}&=-\frac{2h^2(c_b^2-c_f^2)}{c_bc_f^3}
-\frac{\Delta_V}{c_b^3},\nn
\frac{d\lambda}{d\ell}&=\left(\epsilon-\frac{8h^2}{c_f^3}
+\frac{20\Delta_V}{c_b^4}\right)\lambda-\frac{5\lambda^2}{3c_b^3}+\frac{192h^4}{c_f^3},\nn
\frac{dh^2}{d\ell}&=\left(\epsilon-\frac{2\Delta_V}{c_b^4}\right)h^2-\left(\frac{1}{c_f^3}+\frac{8}{c_b(c_b+c_f)^2}\right)4h^4,\nn
\frac{d\Delta_V}{d\ell}&=\left(1+\epsilon-\frac{8h^2}{c_f^3}-\frac{4\lambda}{3c_b^3}\right)\Delta_V+\frac{12\Delta_V^2}{c_b^4},\nonumber
\end{align}
to $\mathcal{O}(\epsilon)$. We find the same clean ($\Delta_V^*=0$) supersymmetric fixed point as before. Linearizing the RG equations around this fixed point, this time we find two relevant directions, one marginal direction, and three irrelevant directions. One of the relevant directions corresponds to the mass parameter $r$ with the same eigenvalue $y_r=2-\epsilon+\mathcal{O}(\epsilon^2)$ as before. The other relevant direction is the only one with a nonzero component along the disorder strength $\Delta_V$, and has the eigenvalue
\begin{align}
y_{\Delta_V}=1-\epsilon+\mathcal{O}(\epsilon^2),\nonumber
\end{align}
which is greater than zero and thus relevant for small $\epsilon$. In fact, all couplings (including the difference between boson and fermion velocities) have a nonzero projection onto this relevant direction. The criterion for the relevance of disorder $y_{\Delta_V}>0$ is equivalent to the Harris criterion, which in the context of the $\epsilon$-expansion should be written as $d<2y_r$ with $d=D-1=3-\epsilon$ and both sides of the inequality are expanded to $\mathcal{O}(\epsilon)$. Although $y_{\Delta_V}$ appears to vanish if $\epsilon$ is naively set to one to reach the physical case of two spatial dimensions, this is most likely only true at linear order in $\epsilon$, and for $\epsilon\rightarrow 1$ corrections of $\mathcal{O}(\epsilon^2)$ and higher cannot be neglected. To the extent that the behavior at small $\epsilon$ is representative of the physical problem, the signatures of the clean quantum critical point with emergent supersymmetry will be visible above a crossover temperature
\begin{align}\label{Tstar}
T^*\sim\Lambda\Delta_{V0}^{1/(1-\epsilon)},
\end{align}
where $\Lambda$ is a high-energy cutoff which for the surface state of a topological insulator can be taken as the bulk energy gap, and $\Delta_{V0}\sim n_0 V^2$ is the bare disorder strength. For temperatures $T<T^*$, the clean quantum critical behavior will be washed out by disorder and the disorder $\Delta_V$ flows to strong coupling. Such a strong disorder fixed point cannot be reliably studied within the present perturbative RG scheme. However, from a general standpoint we propose two possible scenarios. In the first scenario, the effective 4-fermion interaction induced by chemical potential disorder, which is irrelevant in $4-\epsilon$ dimensions at the clean supersymmetric critical point [Eq.~(\ref{eigDmu})], becomes relevant at the strong disorder fixed point. In this case, we expect that a nonzero density of states would be generated for the fermions, and superconductivity would develop as a result of the Cooper instability below a nonzero critical temperature $T_c<T^*$. In that sense, the crossover temperature (\ref{Tstar}) can be seen as an upper bound for $T_c$ at the critical point $g=g_c$ in the disordered system. 

In the second scenario, the chemical potential disorder remains irrelevant at the strong disorder fixed point. The density of states for the Dirac fermions remains zero at $\mu=0$, and $T_c$ is zero at $g=g_c$. The behavior at $g=g_c$ would be controlled by the strong disorder fixed point. This would correspond to a non-monotonic dependence of the superconducting $T_c$ on $g$, where $T_c$ is zero at $g=0$ and $g=g_c$, but nonzero for $0<g<g_c$ and for $g>g_c$. Since such a non-monotonic dependence seems counter-intuitive, we expect this second scenario to be unlikely, and expect that the RG does eventually flow to a superconducting phase. However, we cannot exclude this possibility since the RG flows to strong coupling.

We note that the replica field theoretic analysis assumes translation invariance and neglects mesoscopic fluctuation effects, which were shown to dominate the physics at weak coupling. If the true physics near the clean critical point is also dominated by mesoscopic fluctuations, then the replica field theory approach will dramatically underestimate $T_c$. Conversely, if superconductivity near the clean critical point is spatially uniform, and puddles are unimportant, then the nature of the superconductivity changes between weak coupling and strong coupling. This change in the nature of superconductivity may then be controlled by a strong disorder fixed point. Similar ideas have been discussed for disordered bosons in one dimension in Ref.\onlinecite{Schulz, Altman}. We defer an investigation of these ideas to future work.

\section{Quantum criticality in the disordered system}

It has been pointed out that the quantum critical point in the ideally clean system is described by an unusual effective field theory displaying emergent supersymmetry.\cite{grover, Sung-Sik} However, we have also pointed out that the semimetal phase is itself unstable to disorder. 
Thus, disorder has the effect of destroying the quantum critical point discussed in Ref.~\onlinecite{grover, Sung-Sik} by inducing superconductivity at arbitrarily weak interaction strengths. In this section we discuss to what extent it is possible to observe signatures of the (destroyed) quantum critical point. 

\begin{figure}
\includegraphics[width=\columnwidth]{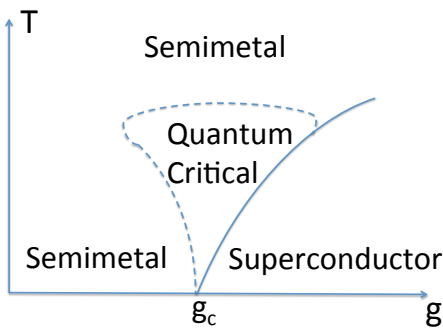}
\caption{Phase diagram of the clean Dirac fermion system at $\mu=0$. The system has only two phases - a superconductor and a semimetal, with a boundary between them which scales as $T_c \sim (g-g_c)^{z \nu}$, where $z = 1$ from Lorentz invariance, and $\nu = \frac{1}{2}+\frac{\epsilon}{4}+\mathcal{O}(\epsilon^2)$ in $4-\epsilon$ spacetime dimensions. Within the semi metallic phase there is a quantum critical regime, which is controlled by the critical point at $g=g_c$. The dashed line indicates a crossover between semi metallic and quantum critical behavior, and is given approximately by Eq.~(\ref{eq: QCboundary}). \label{fig: gtphasediag}}
\end{figure}

We assume that the system has been fine tuned to the Dirac point $\mu=0$. This assumption was also made in Ref.~\onlinecite{grover, Sung-Sik}. The phase diagram of the clean system takes the form Fig.~\ref{fig: gtphasediag}. Note the existence of a `quantum critical regime' at finite temperature. In this regime, one can detect signatures of the proximate quantum critical point. The signatures of the quantum critical point gradually phase out as we move further from the quantum critical point. The dashed line indicates the existence of a crossover between quantum critical and semi-metallic behavior. Note that the quantum critical and semimetallic regions are not different phases. Rather the system evolves smoothly from one to the other. 

Where should one place the boundary of the quantum critical regime? We can answer this question for the clean system as follows. In general, the system may start anywhere in the basin of attraction of the quantum critical point, but it will only start to display quantum critical behavior when all the irrelevant couplings $\lambda_i$ become smaller than some threshold small scale $\lambda_*$. The RG flow equations for the irrelevant couplings $\lambda_i$ take the form
\[
\frac{d \lambda_i}{dl} = - y_{i} \lambda_i,
\]
where the scaling dimensions $y_i$ were calculated in Sec.~V, and $l = \ln \Lambda_0/\Lambda_f$. The RG is started at the initial scale $\Lambda_0 \approx \omega_D$ (below which we have an attractive interaction), and stops at the scale $\Lambda_f \approx T$, where the temperature $T$ supplied the IR cutoff. From this, one obtains the criterion for quantum critical behavior, 
\begin{equation}
T<T_* = \omega_D \min\big(1, (\lambda_*/\lambda^i_0)^{1/y_i} \big), \label{eq: QCboundary}
\end{equation}
where $\lambda_0^i$ is the bare coupling of the $i$th irrelevant operator, $y_i$ is its scaling dimension, and $\lambda_*$ is a small threshold. One should include only those irrelevant couplings which start far away from the critical point $\lambda_0 > \lambda_*$. The precise choice of $\lambda_*$ is somewhat arbitrary. For definiteness, we suggest using $\lambda_* = 0.1$ as a threshold for quantum critical behavior.

\begin{figure}
\includegraphics[width=\columnwidth]{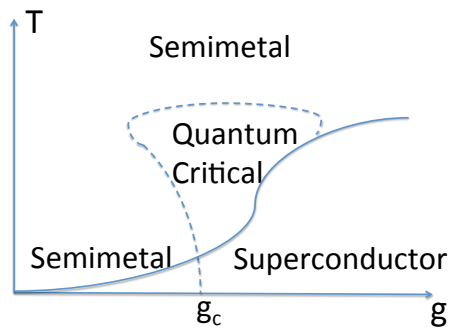}
\caption{Phase diagram of the disordered Dirac system, with very weak disorder. The quantum critical point has been buried under the superconducting phase. However, signatures of the quantum critical point may still be visible in the finite temperature quantum critical regime.\label{fig: dirtyphasediag1}}
\end{figure}

How does the phase diagram change in the presence of disorder? We have argued that in the presence of chemical potential disorder, the semimetal phase is unstable to superconductivity, with a 
critical temperature is given by either (\ref{eq: Josephson1}) or (\ref{eq: TKT}) depending on whether or not the disorder is smooth. Meanwhile, the critical point is also unstable to disorder, although the behavior at $g=g_c$ is controlled by a strong disorder fixed point which we were not able to access in any controlled manner. Although we cannot make definite predictions about the critical theory, since the RG flows to strong disorder, we anticipate that the superconducting $T_c$ should interpolate smoothly between weak and strong coupling. This leads to a phase diagram of the form shown in Fig.\ref{fig: dirtyphasediag1}. Note that the `quantum critical point' has now been buried inside the superconducting phase. 

It is not possible to directly probe the quantum critical point, since it has been buried by the superconducting phase. However, at a temperature above the superconducting critical temperature $T_\textrm{KT}$, one can probe the quantum critical regime, to look for finite temperature signatures of the emergent supersymmetry. Since no real world sample is ever perfectly clean, it follows that signatures of the `emergent supersymmetry' identified in Ref.~\onlinecite{grover, Sung-Sik} can only ever be probed by experiments conducted in this relatively high temperature quantum critical regime. 

\begin{figure}
\includegraphics[width=\columnwidth]{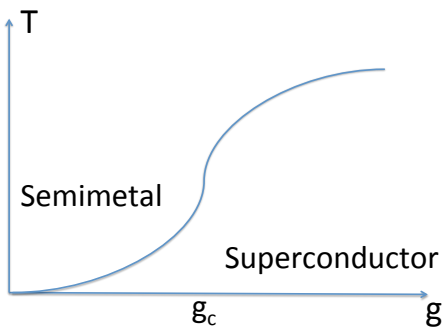}
\caption{Phase diagram of the disordered Dirac system, with less weak disorder. The quantum critical point and the quantum critical regime have both been buried under the superconducting phase. \label{fig: dirtyphasediag2}}
\end{figure}

We note that although the critical temperature for onset of quantum critical physics is of order $\omega_D$ (\ref{eq: QCboundary}), this boundary can be strongly suppressed if the bare theory starts a long way away from the critical point. In principle, it is possible that the quantum critical regime may be entirely buried beneath the superconducting phase, in which case no signatures of the quantum critical point would be detectable in experiments. The resulting phase diagram would then look like (Fig.~\ref{fig: dirtyphasediag2}). This scenario would arise if the critical temperature for quantum critical behavior (\ref{eq: QCboundary}) were less than the critical temperature for superconductivity. 

\section{Conclusions}
We have calculated the phase diagram for a single species of Dirac fermions with attractive delta function interactions. We have shown that the zero temperature phase diagram in the coupling-doping ($g-\mu$) plane consists of a single superconducting phase, except for a line segment along $\mu=0$ and $g<g_c$, which is a semimetal. We have shown that the introduction of disorder destroys the semimetal phase, introducing a finite density of states and triggering onset of superconductivity. Remarkably, disorder actually induces superconductivity, by introducing a non vanishing density of states. This is a striking departure from the usual scenario, where disorder suppresses superconductivity. We note that such a disorder-enhancement of superconductivity has already been observed in numerical simulations.\cite{Franz2012}

The critical temperature at which disorder enhanced superconductivity develops is given by Eq.~(\ref{eq: tcsmoothdisorder}) if the disorder is smooth. The case of short range disorder is more complex. A naive estimate based on a uniform mean field solution for superconductivity gives an estimate (\ref{eq: TcAG}), which is doubly exponentially small in weak disorder. However, superconductivity is strongly enhanced by mesoscopic fluctuation effects, such that the true critical temperature is actually given by Eq.~(\ref{eq: TKT}), which is only exponentially small in weak disorder. The superconducting phase consists of locally superconducting islands, which establish global phase coherence through the Josephson coupling between them (Fig.~\ref{fig: puddles}). Application of a sufficiently strong transverse magnetic field destroys the global phase coherence by frustrating the Josephson couplings, driving the system into a gauge glass phase. 

The region near the critical coupling $g=g_c$ was treated within a replica field theoretic approach. While this approach neglects mesoscopic fluctuations, it is useful for understanding the interplay of disorder and interactions in the strongly coupled theory. We find that chemical potential disorder is a dangerous irrelevant perturbation at the critical point, generating (relevant) disorder in the BCS coupling. This drives an RG flow to strong disorder. While we cannot definitely answer what happens at strong disorder, continuity with the weak coupling results suggests that the RG flow ends up in a superconducting phase. As a result, the quantum critical point identified in Ref.~\onlinecite{grover, Sung-Sik} will be buried under a superconducting phase for any non-vanishing value of disorder. However, for sufficiently weak disorder, signatures of the quantum critical point may still be visible in the finite temperature quantum critical regime. 


This work also suggests some promising future directions for research. Our analysis was focused on the problem with a single Dirac fermion species, but graphene, a popular experimental material, actually possesses four species of Dirac fermions. A generalization of the analysis to graphene would be a useful and worthwhile task. Another potential direction of research would be to further develop the analysis of mesoscopic fluctuations presented in Sec.~IV. While highly suggestive, this analysis was based on the assumption that the SCBA expression for the density of states could be applied at small length scales, to obtain a local density of states from a local concentration of impurities, and that this density of states could be inserted into the BCS calculation. A more rigorous treatment of this issue would be an important addition to the present work. Meanwhile, the analysis at the critical point also opens up some avenues for further research. While we determined that the clean critical point is unstable to weak disorder, we found that the RG flowed to strong disorder. Determining what happens at strong disorder is a worthwhile topic for future work. In addition, the replica symmetric RG analysis ignored the effect of mesoscopic fluctuations, which were known to be important at the Gaussian point. Investigating the effect of mesoscopic fluctuations at the critical point is another topic for future work. Finally, an experimental investigation of the ideas outlined in this paper would present an excellent opportunity to compare theory with experiment. Given the rich phenomenology associated with superconductivity and disorder in Dirac fermion systems, and given the popularity of topological insulators as experimental materials, we urge experimentalists to seach for superconducting topological insulators, and eagerly anticipate further developments in this field. 

Note added: After completion of this work, we became aware of Ref.\onlinecite{Foster2012}. This work looked at the surface states of a topological superconductor with spin SU(2) symmetry, and concluded that they were unstable in the presence of vector potential disorder and interactions. It complements our present work, which looks at the interplay of scalar potential disorder and interactions on the surface states of a topological insulator.  

\acknowledgements

We would like to thank Boris Spivak, Chris Laumann, Liang Fu, Ashvin Vishwanath, and Tarun Grover for useful discussions. This research was supported in part by the National Science Foundation under Grant No.  DMR 08-19860 (DAH) and DMR 10-06608 (SLS) and by the Simons Foundation (JM).

\bibliography{DisorderedDiracSCv2}
\end{document}